\documentclass[pre,twocolumn,showpacs]{revtex4-1}
\usepackage{amsmath,amssymb}
\usepackage{graphicx}
\usepackage{epsfig}

\begin{document}
\title{Inverse transitions in a spin-glass model on a scale-free network}
\author{Do-Hyun Kim}

\affiliation{Jesuit Community, Sogang University, 35 Baekbeom-ro, Mapo-gu, Seoul 121-742, Korea}

\begin{abstract}
In this paper, we will investigate critical phenomena by considering a model spin-glass on scale-free networks. For this purpose, we consider the Ghatak-Sherrington (GS) model, a spin-1 spin-glass model with a crystal field, instead of the usual Ising-type model. Scale-free networks on which the GS model is placed are constructed from the static model, in which the number of vertices is fixed from the beginning.
On the basis of the replica-symmetric solution, we obtain the analytical solutions, i.e., free energy and order parameters, and we derive the various phase diagrams consisting of the paramagnetic, ferromagnetic, and spin glass phases as functions of temperature $T$, the degree exponent $\lambda$, the mean degree $K$, and the fraction of the ferromagnetic interactions $\rho$.
Since the present model is based on the GS model, which considers the three states ($S=0, \pm 1$), the $S=0$ state plays a crucial role in the $\lambda$-dependent critical behavior: glass transition temperature $T_{g}$ has a finite value, even when $2 < \lambda < 3$.
In addition, when the crystal field becomes nonzero, the present model clearly exhibits three types of inverse transitions, which occur when an ordered phase is more entropic than a disordered one.
\end{abstract}


\maketitle

\section{Introduction}

Over the past 10 years, physicists have made a significant effort to understand unusual critical phenomena in complex network systems \cite{Dorogovtsev08}.
Most of the real-world complex networks have been revealed as scale-free (SF)  in the degree
distribution, $P(k) \sim k^{-\lambda}$, where degree $k$ is
the number of edges connected to a given vertex and $\lambda$ is
the degree exponent \cite{Barabasi99,Albert02,Dorogovtsev02a,Newman03}.
The critical phenomena of spin systems on such SF networks have been expected differ from those in Euclidean space.
According to various studies \cite{Aleksiejuk01,Dorogovtsev02b,Leone02,Bianconi02,Herrero04,Suchecki06,Hasegawa07,Iannone11}, the critical behavior of the ferromagnetic Ising model on SF networks has
a strong dependence on the degree distribution.
A spin-glass model has also demonstrated anomalous critical phenomena on SF networks.
The previous analytical study of the Ising spin-glass model on SF networks \cite{Kim05} showed that the phase diagrams
consisting of the paramagnetic (PM), ferromagnetic (FM), and spin-glass (SG) phases were strongly dependent on the given degree
distribution, and the transition temperature $T_{g}$ ($T_{c}$) between the PM-SG (PM-FM) phases became infinite for $2 < \lambda < 3$.

Then, provided an SG model based on the non-Ising type is now considered in SF networks, how will its critical phenomena be changed? Until now, little research has been conducted to answer this interesting question. The main purpose of this paper is to investigate critical phenomena in SF networks by introducing a new non-Ising-type SG model. For this purpose, we consider the Ghatak-Sherrington (GS) model \cite{GS,Costa94}, a spin-1 SG
model with a crystal field. Recently, the GS model has become well known as a
prototypical model for inverse transition \cite{Schupper04,Crisanti05,Leuzzi06,Magalhaes08,Morais12}.

Inverse transition (melting or freezing) occurs when an ordered phase is more entropic than a
disordered one. As a result, the ordered phase appears at
a higher temperature than the disordered one. Such inverse
transitions have been experimentally observed in various
physical systems, such as polymers \cite{Rastogi99,Chevillard97}, high-$T_{c}$ superconductors \cite{Avraham01},
magnetic thin films \cite{Portmann03}, and organic monolayers \cite{Scholl10}. The GS model in Euclidean space shows inverse freezing, wherein the SG phase becomes the one with higher entropy.

Nowadays, the GS and analogous models with a crystal field have drawn attention by the observation of a ``Bose glass" of field-induced magnetic
quasiparticles in a doped quantum magnet \cite{Zapf06,Yu12}. In addition, Erichsen \textit{et al.} obtained the analytical solutions of the GS model with finite connectivity and found inverse transitions \cite{Erichsen11,Erichsen13} by extending the techniques of the previous Ising spin-glass model with finite connectivity \cite{Wemmenhove04}. Meanwhile, a quantum version of the GS model with a transverse tunneling field exhibited a peculiar phenomenon of the splitting within one SG phase, depending on the values of crystal and transverse fields \cite{Kim13}.

In this work, we will study the GS-based SG model on SF networks in order to obtain its analytical solutions, i.e., free energy and order parameters, and various phase diagrams. From these solutions, we can expect two simultaneous unique features when an alternative SG model, based on the GS model, is studied in SF networks.
One such feature is that the alternative model may contain an inverse transition similar to that of the GS model in Euclidean space. The other is that phase diagrams and transition temperatures obtained from the alternative model may exhibit a new dependence of the degree distribution of the considered SF network. However, the divergence of transition temperatures, shown in the Ising spin-glass model on SF networks with $2 < \lambda < 3$, may disappear in the alternative model under the same degree distributions. The following analysis on the alternative model will give clear answers about such expectations.

\section{Model}

Now, we will investigate the critical phenomena by considering the GS model on SF networks. The SF networks, on which the GS model is placed, are constructed by the static model. Based on the replica-symmetric solution, we thereby obtain the analytical solutions, i.e., free energy and order parameters.

\subsection{Construction of `static' SF networks}

In order to construct SF networks, on which spin system is placed, we follow the process of the static model, where the name ``static" originates from the fact that the number of vertices is fixed from the beginning \cite{Goh01}. This model has the advantage for analytical calculations of its theoretical quantities, such as free energy and order parameters \cite{Kim05,Lee04}.

The general random graph under the static model is constructed as follows \cite{Kim05} : The number of vertices $N$ is fixed at the beginning. Each vertex $i$ $(i=1,2,\ldots, N)$ is assigned a weight $p_{i}$. A pair of vertices $(i, j)$ is chosen with the probabilities $p_{i}$ and $p_{j}$, respectively, and they are connected with an edge, unless the pair is already connected. This process is repeated $NK/2$ times. In such random networks, the
probability that a given pair of vertices $(i, j)$ ($i \neq j$) is not connected by an edge, as denoted by $1-f_{ij}$, is given by
$(1-2p_{i}p_{j})^{NK/2} \simeq \exp (-NK p_{i}p_{j})$, while the connection
probability $f_{ij}=1-\exp (-N K p_{i}p_{j})$.

For the Erd\H{o}s-R\'enyi (ER) graph \cite{Erdos59,Bollobas85}, the weight is given as $p_{i}=1/N$,
independent of the index $i$. Since $p_{i} p_{j}=1/N^{2}$, the fraction of bonds present becomes
$f_{ij} \approx K/N$ and the average number of the connected edges is $NK/2$. So $K$ becomes the mean degree in the ER graph.

For the static SF network, the weights are given by
\begin{equation}
p_{i}=\frac{i^{-\mu}}{\zeta_{N}(\mu)}
\end{equation}
where $\mu$ is a control parameter in the range $[0,1)$, and $\zeta_{N}(\mu) \equiv \sum_{j=1}^{N} j^{-\mu} \approx {N^{1-\mu}}/(1-\mu)$. Then the resulting network is an SF network with a power-law degree distribution, $P(k) \sim k^{-\lambda}$, with $\lambda = 1 + 1/\mu$ \cite{Kim05,Goh01}.
Note that $f_{ij} \approx N K p_{i} p_{j}$ for  finite $K$, except $f_{ij} \approx 1$ for $2 < \lambda < 3$ and $ij \ll N^{3-\lambda}$.
Therefore, the mean degree of a vertex $i$ is $N K p_{i}$ and the mean degree of the network is $K$ \cite{Kim05}. When $K$ approaches $N$, this network becomes a fully-connected one or a regular lattice with infinite-range interaction. Therefore, we can say that the mean degree $K$ in this model plays a role similar to the physical dimension for spin systems on lattices.

\subsection{Spin-glass model on SF networks}

Now we consider an alternative SG model on static SF networks. The SG transitions in Euclidean space have already been studied by means of various theoretical methods \cite{Nishimori}. Most of such studies have concentrated on regular lattices or the
infinite-range interaction model on fully connected graphs. To study the SG transitions on the static SF networks, we follow the
previous approach of the dilute Ising SG model with
infinite-range interactions, i.e., the Ising SG model on the ER graph, first performed by Viana and Bray
\cite{Viana85,Kanter87,Mezard87,Mottishaw87,Wong88,Monasson98}, and applied successfully to the Ising SG model on the static SF networks \cite{Kim05} .

The Hamiltonian of the GS model on a graph $G$ constructed by the static model is given as
\begin{eqnarray}
\mathcal{H} = - \sum_{(i,j) \in G} J_{ij} S_{i} S_{j} + D \sum_{i} S_{i}^{2} ~~(S_{i}= 0, \pm 1),
\end{eqnarray}
where $J_{ij}$ is nonzero only when the vertices $i$ and $j$ are connected in $G$.
For the static model, the probability of $G$ in the quenched random network ensemble is defined as
\begin{equation}
P_{K}(G) = \prod_{(i,j) \in G} f_{ij} \prod_{(i,j) \notin G} (1  -f_{ij})
\end{equation}
with $f_{ij} = 1 - \exp (-N K p_{i} p_{j})$, $p_{i}$ being given in Eq.(1).
Then, the ensemble average for a given physical quantity $A$ is taken as
\begin{equation}
\langle A \rangle_{K} = \sum_{G} P_{K}(G) A(G) ,
\end{equation}
where $\langle \cdots \rangle_{K}$ denotes the average over different graph configurations.
In the SG problem, the coupling strengths $\{J_{ij}\}$ are also quenched random variables. We assume that
each $J_{ij}$ is given as $+1$ or $-1$ with probabilities $\rho$
and $1-\rho$, respectively, so the coupling strength distribution is given as
\begin{equation}
P_{\rho}(\{J_{ij}\})=\prod_{(i,j) \in G} \Big[ \rho \delta (J_{ij} - 1) + (1 - \rho) \delta (J_{ij} + 1) \Big].
\end{equation}
The case of $\rho = 1/2$ ($\rho = 1$) corresponds to the fully frustrated (purely ferromagnetic) case, and we consider $\rho$ in the range of $1/2 \le \rho
\le 1$ throughout this work.
Then the ensemble average for a given physical quantity, $A$, is taken as
\begin{equation}
\langle A \rangle_{\rho} = \int dJ_{ij} P_{\rho}(\{J_{ij}\}) A(\{J_{ij}\}) ,
\end{equation}
where $\langle \cdots \rangle_{\rho}$ is an average over the quenched disorder of $J_{ij}$.
Thus, the free energy $F$ is evaluated as $-\beta F =
\langle \langle \ln Z \rangle_{\rho} \rangle_{K}$, where $Z$ is the
partition function for a given distribution of $\{J_{ij}\}$ on a
particular graph $G$.

Here, the replica method is used to evaluate the free
energy, i.e., $-\beta F = \lim_{n \to 0} [\langle \langle Z^{n}
\rangle_{\rho} \rangle_{K}-1]/n$. To proceed, we evaluate the $n$-th
power of the partition function $Z^{n}$,
\begin{widetext}
\begin{eqnarray}
\langle \langle Z^{n} \rangle_{\rho} \rangle_{K} &=&
\textrm{Tr}_{\{S^{\alpha}\}} \Big\langle \Big\langle \exp \Big(\beta
\sum_{ (ij) \in G} J_{ij} \sum_{\alpha}
S_{i}^{\alpha} S_{j}^{\alpha} - \beta D \sum_{i} \sum_{\alpha} (S_{i}^{\alpha})^{2} \Big)
\Big\rangle_{\rho} \Big\rangle_{K} \nonumber \\
&=& \textrm{Tr}_{\{S^{\alpha}\}} \exp \Big(- \beta D \sum_{i} \sum_{\alpha} (S_{i}^{\alpha})^{2} \Big)
\cdot \exp \Big[ \sum_{i < j}\ln
\Big\{ 1 + f_{ij} \Big( \Big\langle \exp (\beta J_{ij}
\sum_{\alpha} S_{i}^{\alpha} S_{j}^{\alpha})
\Big\rangle_{\rho} -1 \Big) \Big\} \Big], \label{z_n}
\end{eqnarray}
where the trace $\textrm{Tr}_{\{S^{\alpha}\}}$ is taken over all
replicated spins $S_{i}^{\alpha} (= 0, \pm 1)$, $\alpha=1,\ldots,n$ is
the replica index, and $\beta=1/T$.
By using the relation,
\begin{eqnarray}
\Big\langle \exp(\beta J_{ij}\sum_{\alpha} S_{i}^{\alpha}
S_{j}^{\alpha}) \Big\rangle_{\rho} = \Big\langle \prod_{\alpha} \Big[
\big( \delta(S_{i}^{\alpha} S_{j}^{\alpha} - 1) + \delta(S_{i}^{\alpha} S_{j}^{\alpha} + 1) \big)\cosh(\beta J_{ij}) 
\big(1 + S_{i}^{\alpha} S_{j}^{\alpha} \tanh(\beta
J_{ij}) \big)  + \delta(S_{i}^{\alpha} S_{j}^{\alpha}) \Big] \Big\rangle_{\rho},~
\end{eqnarray}
applying the Hubbard-Stratonovich identity, and taking the method of steepest descent in the
thermodynamic limit ($N \to \infty$),
the free energy becomes
\begin{eqnarray}
\beta F &=& \frac{1}{2} N K \mathbf{T}_{1}
\sum_{\alpha} q_{\alpha}^{2} + \frac{1}{2} N K \mathbf{T}_{2}
\Big( \sum_{\alpha} q_{\alpha \alpha}^{2} + \sum_{\alpha < \beta} q_{\alpha \beta}^{2}\Big)  + \frac{1}{2} N K \mathbf{T}_{3} \Big(\sum_{\alpha} q_{\alpha \alpha \alpha}^{2}  + \sum_{\alpha < \beta} q_{\alpha \alpha \beta}^{2}
+ \sum_{\alpha < \beta < \gamma} q_{\alpha \beta
\gamma}^{2}\Big) + \cdots \nonumber\\ &&  - \sum_{i} \ln
\textrm{Tr}_{\{S_i^{\alpha}\}} \exp \Big(X_{i} - \beta D\sum_{\alpha} (S_{i}^{\alpha})^{2}\Big),
\end{eqnarray}
where
\begin{eqnarray}
X_{i} &=& N K \mathbf{T}_{1} p_i \sum_{\alpha}
q_{\alpha}S_{i}^{\alpha} +  N K \mathbf{T}_{2} p_i \Big(\sum_{\alpha} q_{\alpha \alpha} (S_{i}^{\alpha})^{2} + \sum_{\alpha <
\beta} q_{\alpha \beta} S_{i}^{\alpha} S_{i}^{\beta} \Big) \nonumber\\
&& + N K \mathbf{T}_{3} p_i \Big( \sum_{\alpha} q_{\alpha \alpha
\alpha} (S_{i}^{\alpha})^{3} + \sum_{\alpha < \beta} q_{\alpha \alpha \beta} (S_{i}^{\alpha})^{2} S_{i}^{\beta}
+ \sum_{\alpha < \beta < \gamma} q_{\alpha \beta
\gamma} S_{i}^{\alpha} S_{i}^{\beta} S_{i}^{\gamma} \Big) +\cdots,
\end{eqnarray}
and
\begin{eqnarray}
\mathbf{T}_{l}(T) \equiv \langle \cosh^n \beta J_{ij} \tanh^{l}
\beta J_{ij} \rangle_{\rho} \stackrel{n \to 0}{\longrightarrow}
[\rho + (-1)^{l}(1-\rho)] \tanh^{l} \beta  ~~~~(l=1,2,\ldots).
\end{eqnarray}
Here $\textrm{Tr}_{\{S_{i}^{\alpha}\}}$ is the trace over the replicated
spins at vertex $i$.
The elements of set $\{\mathbf{q} \}$, i.e., $q_{\alpha}$, $q_{\alpha
\alpha}$, $q_{\alpha \beta}$, $q_{\alpha \alpha \alpha}$, $q_{\alpha \alpha
\beta}$, $q_{\alpha \beta \gamma}$, etc., defined as
\begin{eqnarray}
&& q_{\alpha}=\sum_{i} p_{i} \langle S_{i}^{\alpha}\rangle_{i}, ~~
q_{\alpha \alpha}=\sum_{i} p_{i} \langle (S_{i}^{\alpha})^{2} \rangle_{i}, ~~
q_{\alpha \beta}=\sum_{i} p_{i} \langle S_{i}^{\alpha} S_{i}^{\beta} \rangle_{i}, ~~ \nonumber\\
&& q_{\alpha \alpha \alpha}=\sum_{i} p_{i} \langle (S_{i}^{\alpha})^{3} \rangle_{i},~~
q_{\alpha \alpha \beta}=\sum_{i} p_{i} \langle (S_{i}^{\alpha})^{2} S_{i}^{\beta} \rangle_{i},~~
q_{\alpha \beta \gamma}=\sum_{i} p_{i} \langle S_{i}^{\alpha} S_{i}^{\beta} S_{i}^{\gamma}\rangle_{i},
\cdots,
\end{eqnarray}
\end{widetext}
are the order parameters of the spin glass system, called the
magnetization, the spin self-interaction, the spin-glass order parameter, and so on.
The average is evaluated through $\langle A \rangle_{i} \equiv
\textrm{Tr}_{\{S_{i}^{\alpha}\}} A \exp
X_{i}/\textrm{Tr}_{\{S_{i}^{\alpha}\}} \exp X_{i}$.

\subsection{Replica-symmetric solutions}

Now we consider the replica-symmetric (RS) case, in which spins with
different replica indices are indistinguishable, for simplicity.
Furthermore, since order parameters such as $q_{\alpha \beta \gamma \delta}$ are too complex to be obtained, for the present, we follow an
approach that is similar in spirit to the Sherrington-Kirkpatrick model \cite{SK},  in which terms that are higher in order than
$q_{\alpha \beta}$ in Eqs.(9) and (10) are neglected.
We thus determine the phase boundaries of the PM, FM, and SG phases through the RS solution with three order parameters: the RS
magnetization, the RS spin self-interaction, and the RS SG order parameter are denoted as
$M(=q_{\alpha})$,  $R(=q_{\alpha \alpha})$, and $Q(=q_{\alpha \beta})$, respectively, and the
free energy in Eq.(9) is truncated at the
order of $q$. Then we obtain the RS intensive free energy as follows:
\begin{widetext}
\begin{equation}
\beta f(M, R, Q) = \frac{1}{4} K \mathbf{T}_{2} (R^{2} - Q^{2}) + \frac{1}{2} K \mathbf{T}_{1} M^{2} - \int \mathcal{D}z \frac{1}{N} \sum_{i}
\ln \Big[ 1+ 2 e^{\gamma_{i}} \cosh \eta_{i}(z) \Big].
\end{equation}
\end{widetext}
where $\int \mathcal{D}z \cdots \equiv \frac{1}{\sqrt{2\pi}}
\int_{-\infty}^{\infty} dz ~e^{-z^2/2} \cdots$, $\gamma_{i} \equiv \frac{1}{2} N K \mathbf{T}_{2}p_{i} (R-Q) - \beta D$ and $\eta_{i}(z)
\equiv z \sqrt{N K \mathbf{T}_{2} p_{i} Q} + N K \mathbf{T}_{1} p_{i} M$.

We can determine  $M$, $R$, and $Q$ by imposing the condition that $f$
resumes the stable extrema when they are the replica-symmetric
solutions. From this extremal condition, we
can obtain the self-consistent equations of $M$, $R$, and $Q$
(i.e., $\partial f/\partial M = \partial f/\partial R = \partial f/\partial Q = 0$) as follows:
\begin{eqnarray}
M &=& \int \mathcal{D}z~ \sum_{i} p_{i}\Bigg[\frac{2e^{\gamma_{i}}
\sinh \eta_{i}(z)}{1+2e^{\gamma_{i}} \cosh \eta_{i}(z)}\Bigg]\\
R &=& \int \mathcal{D}z~ \sum_{i} p_{i}\Bigg[\frac{2e^{\gamma_{i}}
\cosh \eta_{i}(z)}{1+2e^{\gamma_{i}} \cosh \eta_{i}(z)}\Bigg]\\
Q &=& \int \mathcal{D}z~ \sum_{i} p_{i}\Bigg[\frac{2e^{\gamma_{i}}
\sinh \eta_{i}(z)}{1+2e^{\gamma_{i}} \cosh \eta_{i}(z)} \Bigg]^{2}
\end{eqnarray}

Now, we consider three phases: PM ($M=Q=0$), SG ($M=0$, $Q \neq 0$), and FM ($M \neq 0$, $Q \neq 0$).
Under certain conditions, there exists a multicritical point at which the PM-SG-FM phases merge.
In addition to the PM, SG, and FM phases, the mixed (M) phase is sometimes present in the present model.
The M phase is defined as the re-entrant SG phase with nonzero
 ferromagnetic order ($M \neq 0$, $Q \neq 0$), located below the FM phase \cite{Fischer91,Mydosh93}. The SG-M phase
boundary is determined as the vertical straight line (Toulouse line) from the
multicritical point to $T/J=0$ \cite{Toulouse80}. The phase boundary between the replica symmetric phase (PM, FM)
and the replica-symmetry-broken phase (SG, M) is determined by using the Almeida-Thouless (AT)
line \cite{Almeida78},
\begin{widetext}
\begin{eqnarray}
[AT] \equiv  (K \mathbf{T}_{2})^{-1} - \int \mathcal{D}z \sum_{i} N p_{i}^2~
\Bigg[\frac{4e^{2\gamma_{i}} (2e^{\gamma_{i}} + \cosh \eta_{i}(z))^{2} }
{(1+2e^{\gamma_{i}} \cosh \eta_{i}(z))^{4}} \Bigg] = 0.
\end{eqnarray}
\end{widetext}
Both the PM-SG phase boundary and the FM-M phase boundary are determined by the AT line. Note that the RS solutions [Eqs. (14)-(16)] are unstable
below the AT line, i.e., inside the SG and M phases.
We can thus complete the phase diagrams of the present model from these equations.
We consider four phases: PM ($M=Q=0$, $[AT]>0$), SG ($M=0$, $Q \neq 0$, $[AT]<0$), FM ($M \neq 0$, $Q \neq 0$, $[AT]>0$), and M ($M \neq 0$, $Q \neq 0$, $[AT]<0$).

Note that the ER-type model can be obtained when we select the weight $p_{i}=1/N$.  Then our results ($f$, $M$, $R$, $Q$, and $[AT]$) become simpler and independent of index $i$. Then the results appear mathematically identical to those of the GS model, except for the following main difference: Instead of $\beta^{2}$ of the GS model, $K \tanh^{2} \beta$ is used in the ER-type model.
The difference may be overcome when $\beta \ll 1$, i.e., $T \gg 1$ and $K \to 1$. Therefore, in $T \ll 1$ and $K \gg 1$, the main region of the inverse transitions, the GS model and the present model have very distinctive characteristics, in spite of the mathematical similarity of the results. Moreover, when $p_{i}$ depends on index $i$ in the SF model, the distinctiveness becomes larger. The distribution of $J_{ij}$ also differs between the two models: While the present model is based on the $\pm J$ model, the GS model has a Gaussian distribution of $J_{ij}$. Therefore, we may not obtain any meaningful information from the simple and direct comparison between the results of these two models.

The present model demonstrates a huge difference from the previous Ising SG model \cite{Kim05}, in that our model has no divergence of $T_{g}$ in the region $2 < \lambda < 3$: for the previous Ising SG model \cite{Kim05}, $M$ and $Q$ approach zero near the PM-SG phase boundary, so the key term comprising free energy, $\ln [2 \cosh \eta_{i}(z)]$ in Eq.(20) of Ref. \cite{Kim05}, can be expanded by a series expansion of $\eta_{i}^{2}(z)$ and $\eta_{i}^{4}(z)$, i.e., the even multiples of small $\eta_{i}(z) (\ll 1)$. According to Ref. \cite{Kim05}, the term $K_{p}$ of the model was introduced by the series expansion, and the divergence of $T_{g}$ in the region $2 < \lambda < 3$ occurred with the introduction of $K_{p}$. However, the present model does not allow for such a series expansion:  $1+2e^{\gamma_{i}} \cosh \eta_{i}(z)$ in Eq.(13) is always larger than 1, although $M$ and $Q$ approach zero near the PM-SG phase boundary. Here, the ``1" of $1+2e^{\gamma_{i}} \cosh \eta_{i}(z)$ results from $S=0$, which is the differentiating spin value of the present model from the previous one. Thus, the present model has no room for the introduction of $K_{p}$. Therefore, this model has no reason for any divergence of $T_{g}$ in the region $2 < \lambda < 3$. $T_{g}$ in $\lambda \to 2.0^{+}$ thus has a finite value, which is the main difference from the result of the previous model.

\subsection{Perturbative approach}

In the previous subsection, we neglected such terms that are higher in order than $q_{\alpha \beta}$ in Eqs.(9) and (10).
Now we consider the perturbative approach \cite{Kim05,Viana85}, by which we expand the term of $\ln
\textrm{Tr}_{\{S_i^{\alpha}\}} \exp \big(X_{i} - \beta D\sum_{\alpha} (S_{i}^{\alpha})^{2}\big)$ in Eq.(9) up to fourth order.
Through this approach, we obtain the order parameters $q_{\alpha}$, $q_{\alpha \alpha}$, $q_{\alpha \beta}$, $q_{\alpha \alpha \alpha}$, $q_{\alpha \alpha \beta}$, $q_{\alpha \beta \gamma}$ and so on. For simplicity, we use the notations defined by
$\mathcal{Q}_{\alpha} \equiv K \mathbf{T}_{1} q_{\alpha}$,
$\mathcal{Q}_{\alpha \alpha} \equiv K \mathbf{T}_{2} q_{\alpha \alpha}$,
$\mathcal{Q}_{\alpha \beta} \equiv K \mathbf{T}_{2} q_{\alpha \beta}$,
$\mathcal{Q}_{\alpha \alpha \alpha} \equiv K \mathbf{T}_{3} q_{\alpha \alpha \alpha}$,
$\mathcal{Q}_{\alpha \alpha \beta} \equiv K \mathbf{T}_{3} q_{\alpha \alpha \beta}$,
$\mathcal{Q}_{\alpha \beta \gamma} \equiv K \mathbf{T}_{3} q_{\alpha \beta \gamma}$,
$\mathcal{Q}_{\alpha \alpha \alpha \alpha} \equiv K \mathbf{T}_{4} q_{\alpha \alpha \alpha \alpha}$,
$\mathcal{Q}_{\alpha \alpha \alpha \beta} \equiv K \mathbf{T}_{4} q_{\alpha \alpha \alpha \beta}$,
$\mathcal{Q}_{\alpha \alpha \beta \beta} \equiv K \mathbf{T}_{4} q_{\alpha \alpha \beta \beta}$,
$\mathcal{Q}_{\alpha \alpha \beta \gamma} \equiv K \mathbf{T}_{4} q_{\alpha \alpha \beta \gamma}$,
and $\mathcal{Q}_{\alpha \beta \gamma \delta} \equiv K \mathbf{T}_{4} q_{\alpha \beta \gamma \delta}$.
Let $O$ represent a subset of the replica indices $\{1,2,\ldots, n \}$. Then
it is convenient to denote the set
$\{\mathcal{Q}_{\alpha}, \mathcal{Q}_{\alpha \alpha}, \mathcal{Q}_{\alpha \beta},\ldots \}$ as $\{\mathcal{Q}_{O}\}$. We also write $\sigma_{O i} \equiv \prod_{\alpha \in O} S_{i}^{\alpha} = 0, \pm 1$. With these notations,
$X_{i} = N p_{i} \sum_{O} \mathcal{Q}_{O} \sigma_{O i}$ where
the sum is over all subsets of $\{1,2,\ldots,n\}$ except the null set, and
\begin{eqnarray}
e^{X_{i}} &=&  \prod_{O} e^{N p_{i} \mathcal{Q}_{O} \sigma_{O i}} \nonumber\\
&=& \prod_{O} \cosh (N p_{i} \mathcal{Q}_{O}) \times \prod_{O}(1+ \tau_{O}
\sigma_{O i})
\end{eqnarray}
where $\tau_{O} \equiv \tanh (N p_{i} \mathcal{Q}_{O})$. This perturbative
approach is to expand $\prod_{O}(1+\tau_{O}\sigma_{O i})$ and keep only the terms up to fourth order.

Using the properties that $\textrm{Tr} \sigma_{O i}=0$,
$\textrm{Tr} \sigma_{O i} \sigma_{O' i}=0$ for $O \ne O'$
and so on, we rewrite the intensive free energy up to fourth order terms as
\begin{eqnarray}
\beta f &=& \frac{1}{2 K \mathbf{T}_{1}} \sum_{\alpha} \mathcal{Q}_{\alpha}^{2}
+ \frac{1}{2 K \mathbf{T}_{2}} \sum_{\alpha} \mathcal{Q}_{\alpha \alpha}^{2}
+ \frac{1}{2 K \mathbf{T}_{2}} \sum_{\alpha < \beta} \mathcal{Q}_{\alpha \beta}^{2} \nonumber\\
&& + \frac{1}{2 K \mathbf{T}_{3}} \sum_{\alpha} \mathcal{Q}_{\alpha \alpha \alpha}^{2}
+ \frac{1}{2 K \mathbf{T}_{3}} \sum_{\alpha < \beta} \mathcal{Q}_{\alpha \alpha \beta}^{2}  \nonumber\\
&& + \frac{1}{2 K \mathbf{T}_{3}} \sum_{\alpha < \beta < \gamma} \mathcal{Q}_{\alpha \beta \gamma}^{2}
+ \frac{1}{2 K \mathbf{T}_{4}} \sum_{\alpha} \mathcal{Q}_{\alpha \alpha \alpha \alpha}^{2}  \nonumber\\
&& + \frac{1}{2 K \mathbf{T}_{4}} \sum_{\alpha < \beta} \mathcal{Q}_{\alpha \alpha \alpha \beta}^{2}
+ \frac{1}{2 K \mathbf{T}_{4}} \sum_{\alpha < \beta} \mathcal{Q}_{\alpha \alpha \beta \beta}^{2} \nonumber\\
&& + \frac{1}{2 K \mathbf{T}_{4}} \sum_{\alpha < \beta < \gamma} \mathcal{Q}_{\alpha \alpha \beta \gamma}^{2}
+ \frac{1}{2 K \mathbf{T}_{4}} \sum_{\alpha < \beta < \gamma < \delta} Q_{\alpha \beta \gamma \delta}^{2} \nonumber\\
&&  -\frac{1}{N}\sum_{i} \ln \bigg[ 1 + e^{- \beta D} \prod_{O} \cosh (N p_{i} \mathcal{Q}_{O}) \nonumber\\
&& \times \Big\{ 1 + \sum_{\alpha < \beta} \tau_{\alpha} \tau_{\beta} \tau_{\alpha \beta} \nonumber\\
&& + \sum_{\alpha < \beta} \tau_{\alpha} \tau_{\beta} \tau_{\alpha \alpha} \tau_{\alpha \beta}
+ \sum_{\alpha < \beta < \gamma} \tau_{\alpha} \tau_{\beta} \tau_{\gamma} \tau_{\alpha \beta \gamma} \nonumber\\
&& + \sum_{\alpha < \beta < \gamma} ( \tau_{\alpha} \tau_{\beta} \tau_{\alpha \gamma} \tau_{\gamma \beta}
+ \tau_{\beta} \tau_{\gamma} \tau_{\beta \alpha} \tau_{\alpha \gamma}  + \tau_{\alpha} \tau_{\gamma} \tau_{\alpha \beta} \tau_{\beta \gamma} ) \nonumber\\
&& + \sum_{\alpha < \beta < \gamma} \tau_{\alpha \beta} \tau_{\beta \gamma} \tau_{\gamma \alpha} \nonumber\\
&& + \sum_{\alpha < \beta} \tau_{\alpha \alpha} \tau_{\alpha\beta} \tau_{\alpha \alpha \alpha \beta}
+ \sum_{\alpha < \beta} \tau_{\alpha \alpha} \tau_{\beta \beta} \tau_{\alpha \alpha \beta \beta} \nonumber\\
&& + \sum_{\alpha < \beta < \gamma} \tau_{\alpha \alpha} \tau_{\beta \gamma} \tau_{\alpha \alpha \beta \gamma}
+ \sum_{\alpha < \beta < \gamma < \delta} \tau_{\alpha} \tau_{\beta} \tau_{\gamma} \tau_{\delta} \tau_{\alpha \beta \gamma \delta} \nonumber\\
&& + \sum_{\alpha < \beta < \gamma < \delta} (\tau_{\alpha \beta} \tau_{\gamma \delta}
+ \tau_{\alpha \gamma} \tau_{\beta \delta} + \tau_{\alpha \delta} \tau_{\beta \gamma}) \tau_{\alpha \beta \gamma \delta} \nonumber\\
&& + \sum_{\alpha < \beta < \gamma < \delta} \tau_{\alpha \beta} \tau_{\beta \gamma} \tau_{\gamma \delta} \tau_{\delta \alpha} \Big\} \bigg].
\end{eqnarray}

In this case, the ``1" of the $\ln [1 + \cdots]$ term in Eq.(19) results from $S = 0$. This simple effect by $S = 0$ makes a huge difference with the previous Ising SG model \cite{Kim05}, as already pointed out in the previous subsection: Provided the ``1" is absent, Eq.(19) becomes similar to  Eq.(38) of Ref. \cite{Kim05}, so order parameters $\mathcal{Q}_{O}$ can show singularities by divergences of $T_{g}$ (or $T_{c}$) depending on the $\lambda$ value.
However, the existence of the ``1" blocks any possibility of their singularity depending on $\lambda$ value. Furthermore, terms having higher order $\mathbf{T}$, such as $\tau_{\alpha \beta \gamma \delta}$, $1/(2 K \mathbf{T}_{3}) \sum \cdots$ and $1/(2 K \mathbf{T}_{4}) \sum \cdots$ terms, in Eq.(19) are smaller than the rest ones, so the contributions by higher order terms such as $\mathcal{Q}_{\alpha \beta \gamma}$ or $\mathcal{Q}_{\alpha \beta \gamma \delta}$ may be negligible compared with those by $\mathcal{Q}_{\alpha}$ and $\mathcal{Q}_{\alpha \beta}$ near the phase transition points in the thermodynamic limit.
Thus, it becomes sufficiently meaningful to consider only the three RS order parameters $M$, $R$, and $Q$ given in Eqs. (14)-(16).
From now onward, we consider the physical properties of the RS solutions and phase diagrams in detail.

\section{Results}

\begin{figure}
\includegraphics[width=0.43\textwidth]{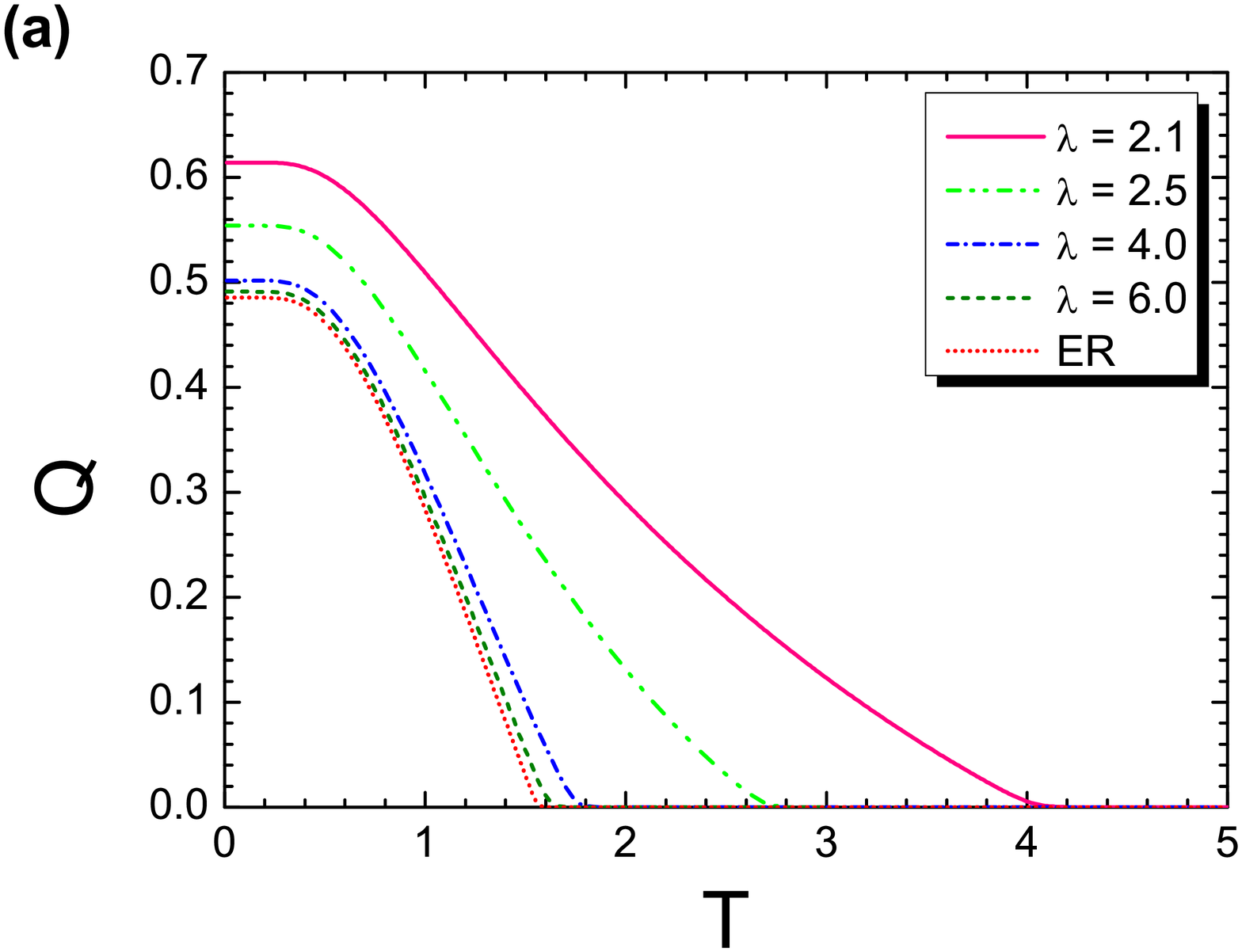}
\includegraphics[width=0.43\textwidth]{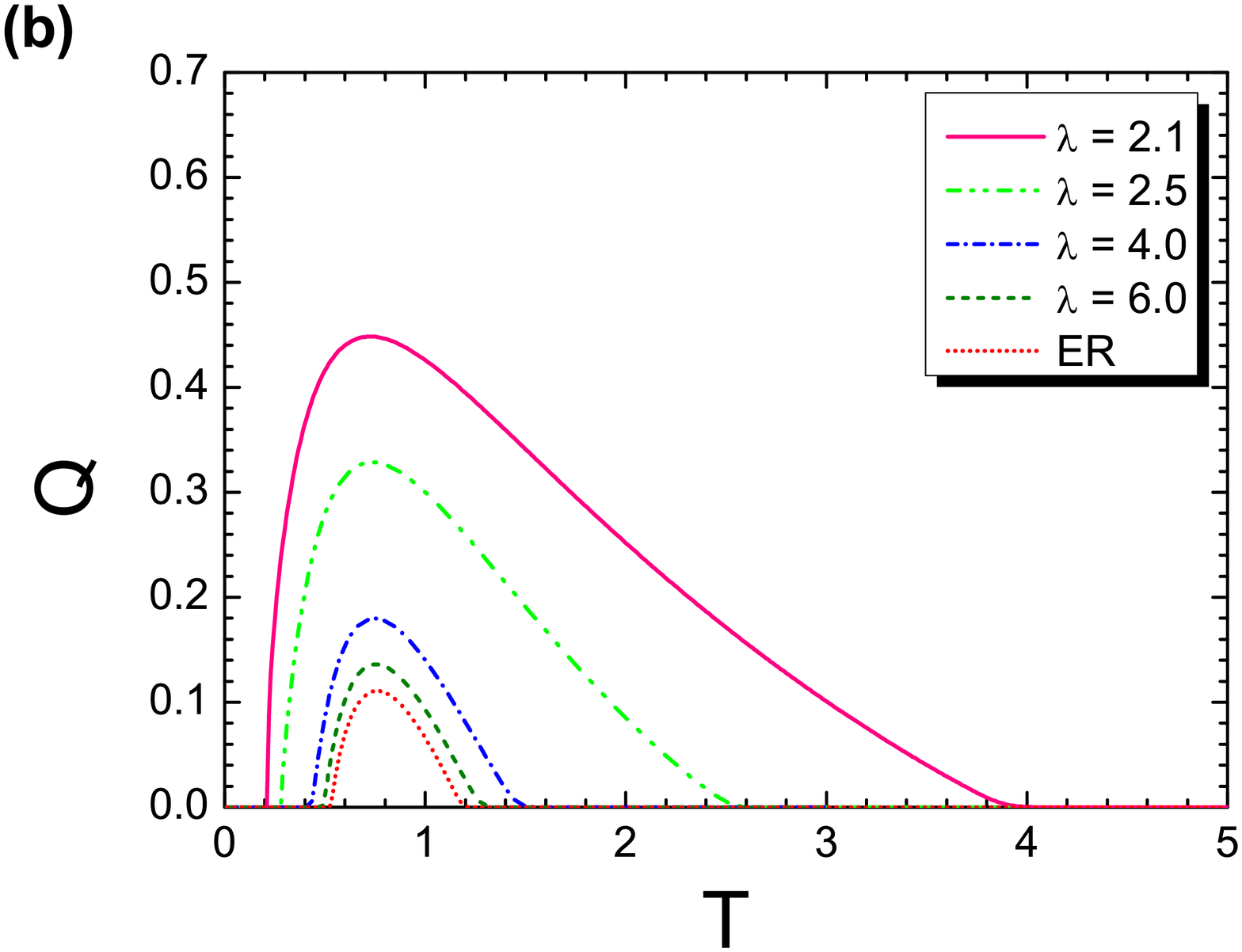}
\includegraphics[width=0.43\textwidth]{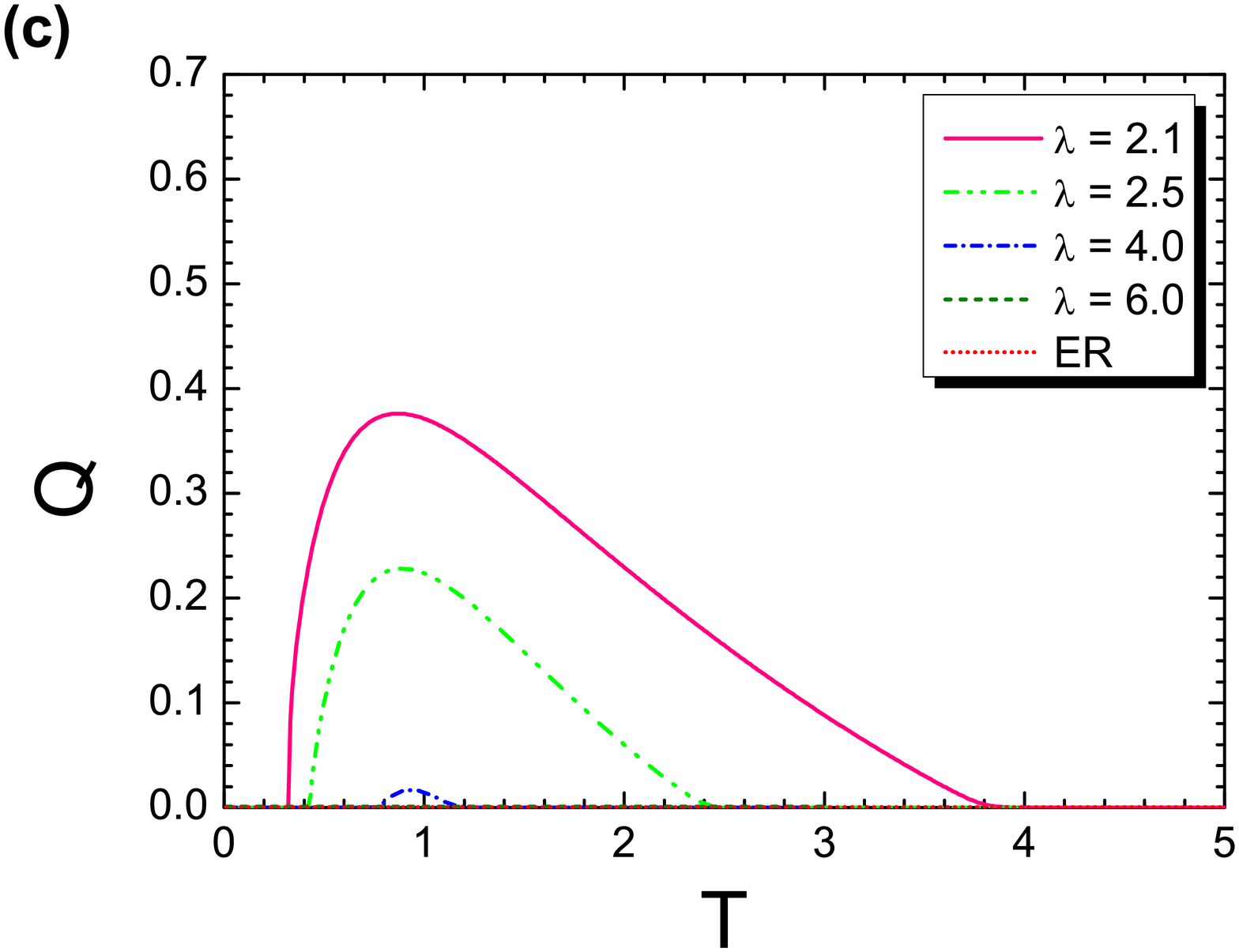}
\caption{(Color online) Values of $Q$ for $D = 0.0$ (a),
$D = 1.0$ (b), and $D = 1.5$ (c). Here, $N = 100$, $\rho = 0.5$, and $K = 5.0$.}\label{1}
\end{figure}

Figure 1 shows the dependence of $Q$  on $D$. $Q$ has a finite value of $T_{g}$, even at $\lambda = 2.1$. As $\lambda$ gradually increases, $Q$ approaches an ER case. The $D=0.0$ case, as shown in Fig. 1(a), shows no inverse freezing. There is only the SG phase at any temperature in the range $0 \leq T \leq T_{g}$. Furthermore, $T_{g}$ approaches a finite value as $\lambda \to 2.0^{+}$.
It is well known that a SF network with a finite $\lambda$ value over 2.0 has a few hubs holding together numerous vertices with small degrees, but the ER network does not have such a hub. Therefore, as $\lambda$ increases gradually from 2.0, the number of hubs inside the SF network also decreases. Since the hubs have a large value of degrees, they tend to be frustrated by spins of numerous neighbor vertices. Since SG is defined as a complex system characterized by frustration, if the hubs under consideration have frustration, they play a crucial role in maintaining SG glass phase against any variation of physical parameters such as $T$ and $D$. As $\lambda$ increases from 2.0, $T_{g}$ thus decreases in proportion to the decrease of hubs.

In Fig. 1(b), an inverse transition occurs for all $\lambda$ values when the $D$ value is nonzero. The inverse transition has two critical temperatures, $T_{g}$ and $T_{p}$, for the second-order P-SG and the first-order SG-P phase transitions.
The phase is varied in the order of PM $\stackrel{\mathrm{2nd}}{\longrightarrow}$ SG $\stackrel{\mathrm{1st}}{\longrightarrow}$ PM, as the temperature is reduced. This inverse transition thus corresponds to inverse freezing, according to the definition of Schupper and Shnerb \cite{Schupper04}. Inverse freezing depends on the value of $\lambda$: as $\lambda$ increases from 2.0, $T_{g}$ decreases but $T_{p}$ increases, because the range of the SG phase becomes narrower. The ER case with no hub has the narrowest region of the SG phase. Note that the phase transition at $T_{g}$ is of second-order, but the transition at $T_{p}$ is a first-order one. As shown in Fig. 1(c), the larger value of $D$ reduces the difference between $T_{g}$ and $T_{p}$. For $\lambda = 6.0$ and for the ER case, no SG phase exists in the entire range of temperatures. The value of $D$ thus has a crucial effect on the inverse freezing of this model. The effect of the $D$ field can be more clearly checked from phase diagrams shown in Fig. 2.

\begin{figure}
\includegraphics[width=0.43\textwidth]{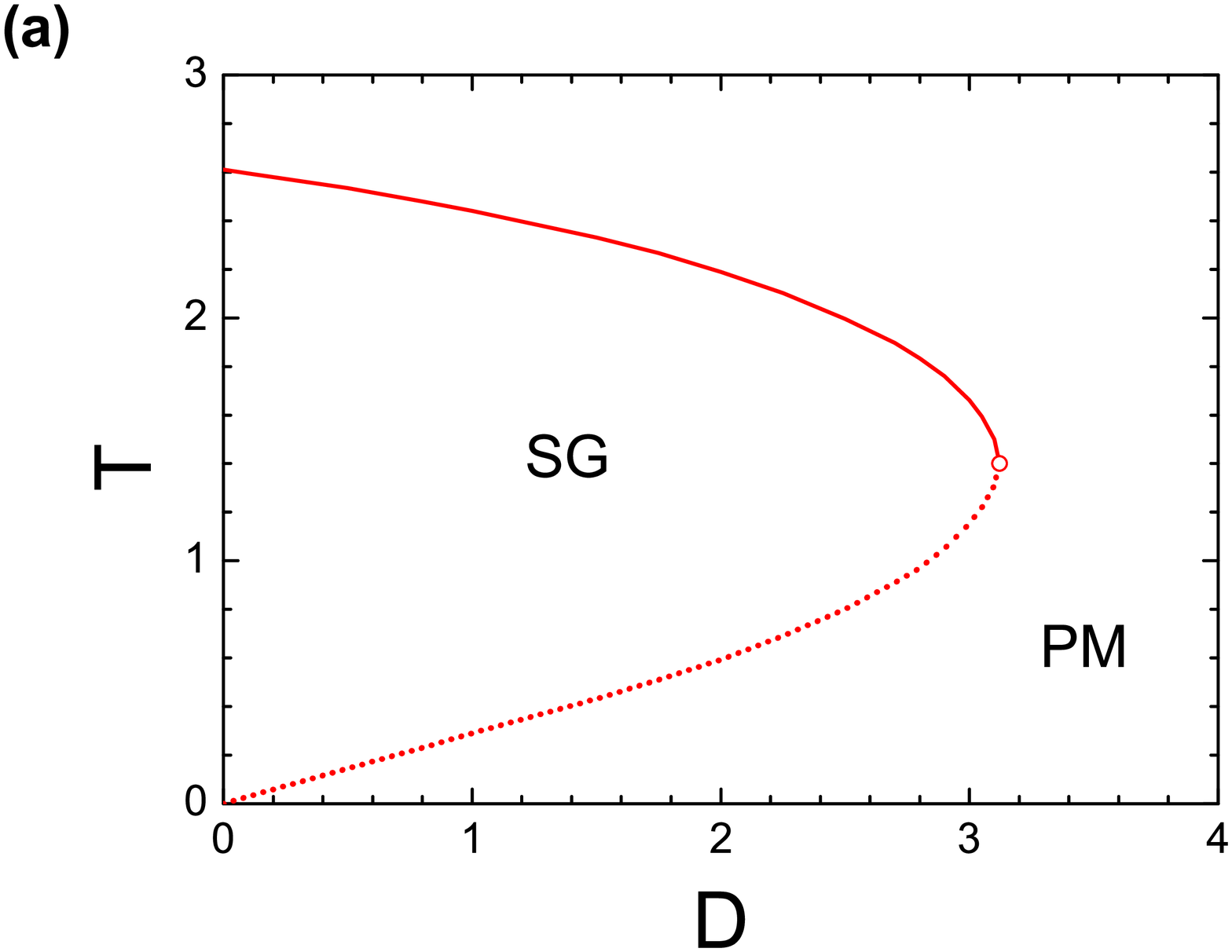}
\includegraphics[width=0.43\textwidth]{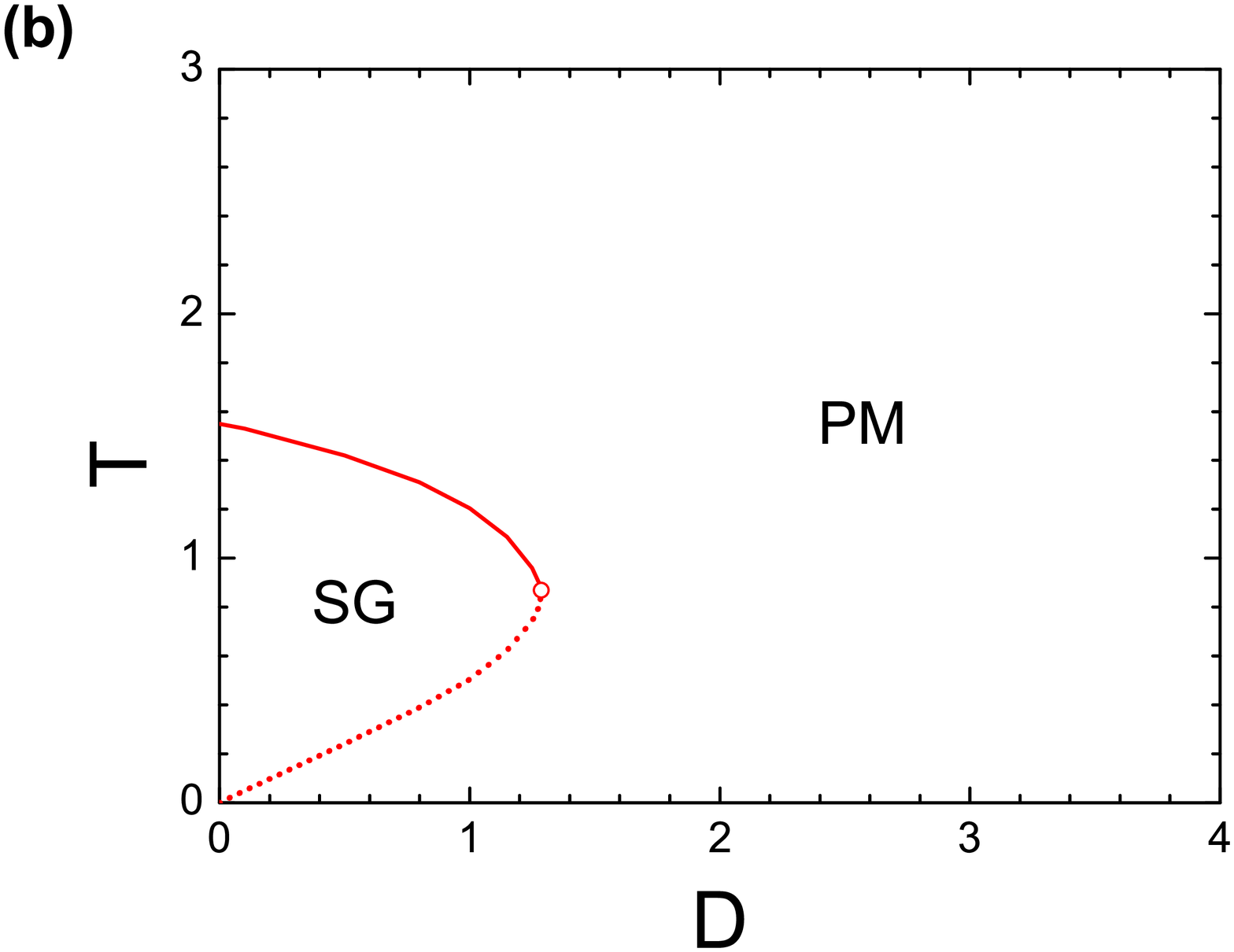}
\caption{(Color online) The $T-D$ phase diagrams for $\lambda=2.5$ (a)  and $\lambda=6.0$ (b).
 Here $N = 100$, $\rho = 0.5$, and $K = 5.0$. The solid-line (dotted-line) part of each phase boundary indicates the second-order (first-order) phase transition and each circle between the two kinds of lines denotes a tricritical point.}\label{2}
\end{figure}

The graphs in Fig. 2 show the $T-D$ phase diagrams obtained for specific $\lambda$ values. For nonzero $D$, inverse freezing occurs as the temperature is lowered. Even a very small value of $D$ can result in inverse freezing, by which the phase varies in the order of PM $\stackrel{\mathrm{2nd}}{\longrightarrow}$ SG $\stackrel{\mathrm{1st}}{\longrightarrow}$ PM, as the temperature is reduced. Figure 2(a) shows a tricritical point (TCP), i.e., the cross-point between first- and second-order phase boundaries, at $D=3.12$. In the region of $D > 3.120$, therefore, only the PM phase exists. In Fig. 2(b) with $\lambda=6.0$, a larger value than that in Fig. 2 (a) causes $T_{g}$ to decrease and $T_{p}$ to increase. A TCP of this case is therefore located at the smaller value, $D=1.286$, than that of Fig. 2(a). The $\lambda$ value thus plays an important role in the variation of $T_{g}$ and $T_{p}$.   Such a role is clearly shown in Fig. 3.

\begin{figure}
\includegraphics[width=0.43\textwidth]{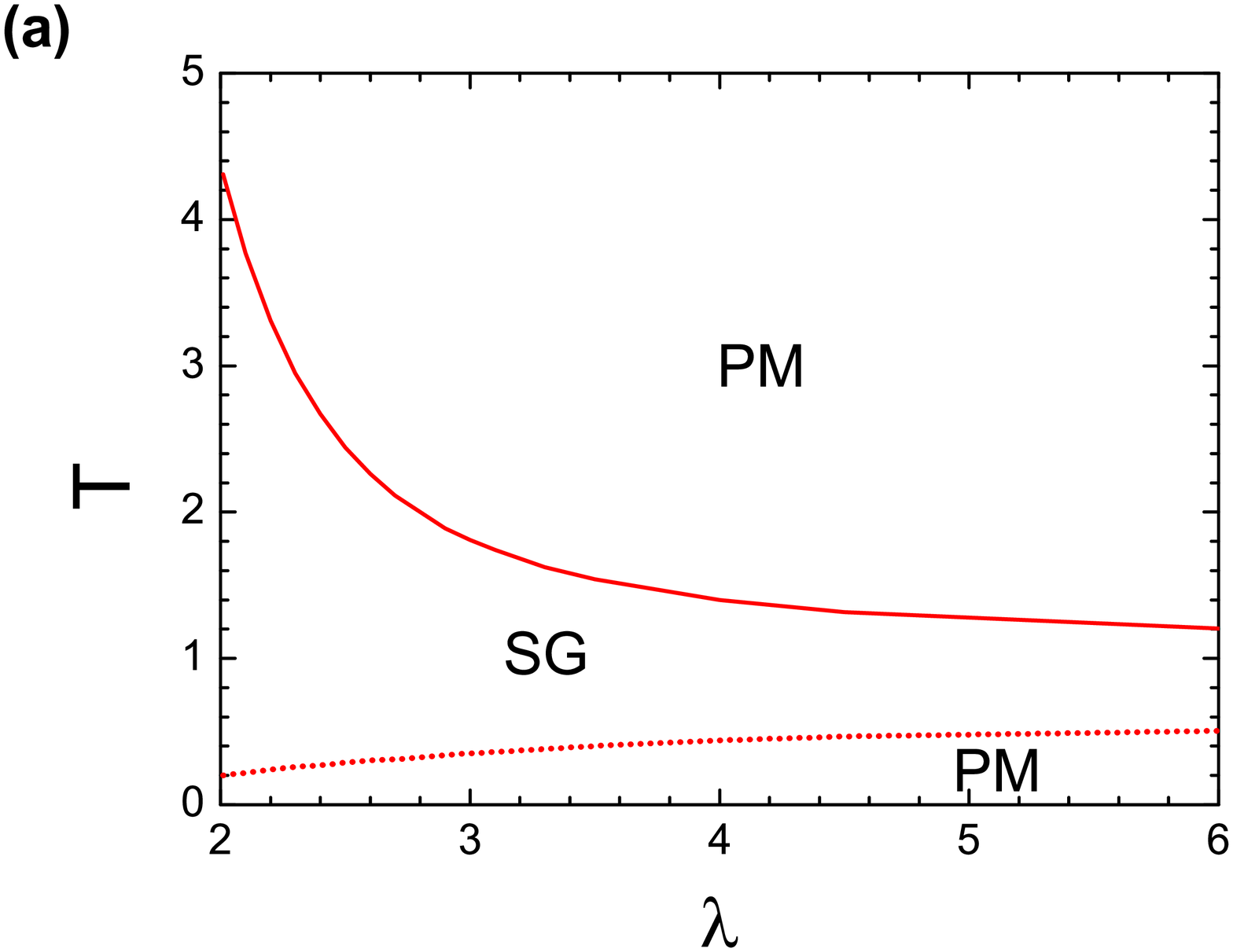}
\includegraphics[width=0.43\textwidth]{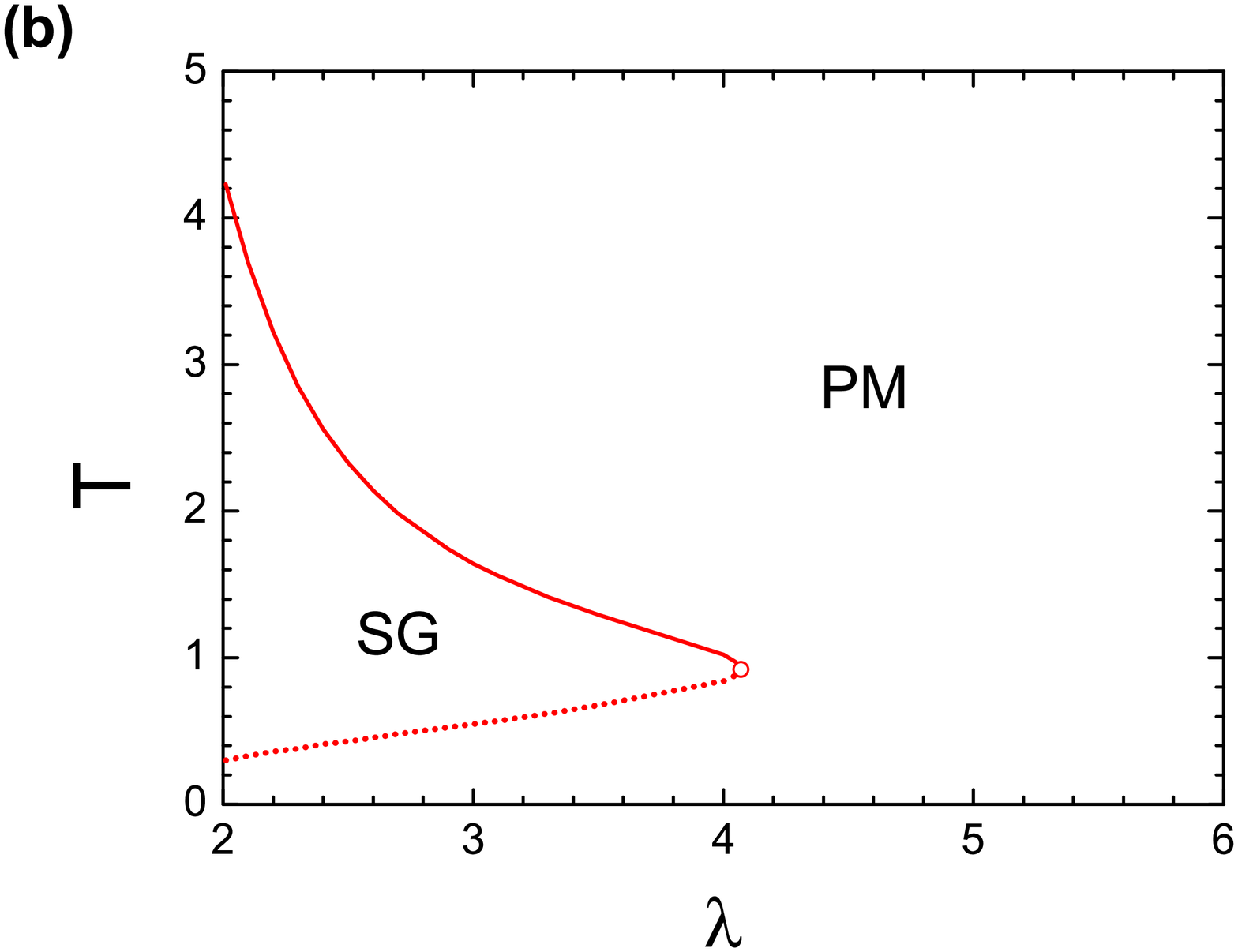}
\caption{(Color online) The $T-\lambda$ phase diagrams for $D=1.0$ (a) and $D=1.5$ (b).
 Here, $N = 100$, $\rho = 0.5$, and $K = 5.0$. }\label{3}
\end{figure}

The phase diagrams in Fig. 3 show the dependence of $T_{g}$ and $T_{p}$ on $\lambda$. As shown in Fig. 3(a), $T_{p}$ and $T_{g}$ under the nonzero $D$ value is also nonzero and finite, respectively, even at $\lambda \to 2.0^{+}$, which was already checked in Fig. 1(b). Figure 3(a) clearly reveals that $T_{g}$ decreases and $T_{p}$ increases as $\lambda$ increases from 2.0. The region of the SG phase thus becomes narrower as $\lambda$ increases, but it does not disappear, even for a large value of $\lambda$. As shown in Fig. 3(b), however, for sufficiently large $D$, the $T-\lambda$ phase diagram has a TCP at a specific $\lambda$ value. The case of $D=1.5$ shows the TCP at $\lambda = 4.070$. It is obvious that as the value of $D$ is increased, the TCP is located at a smaller value of $\lambda$.

\begin{figure}
\includegraphics[width=0.43\textwidth]{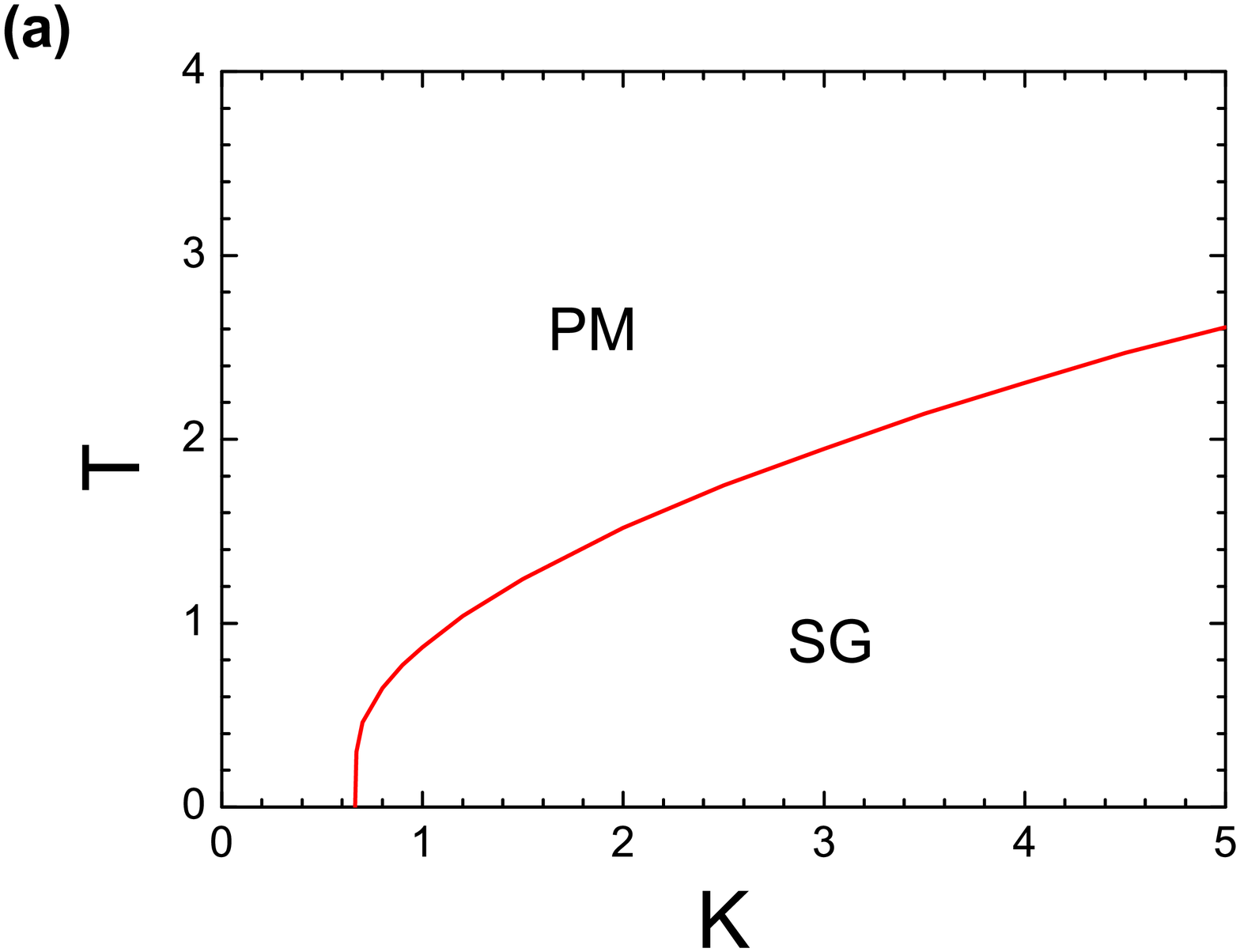}
\includegraphics[width=0.43\textwidth]{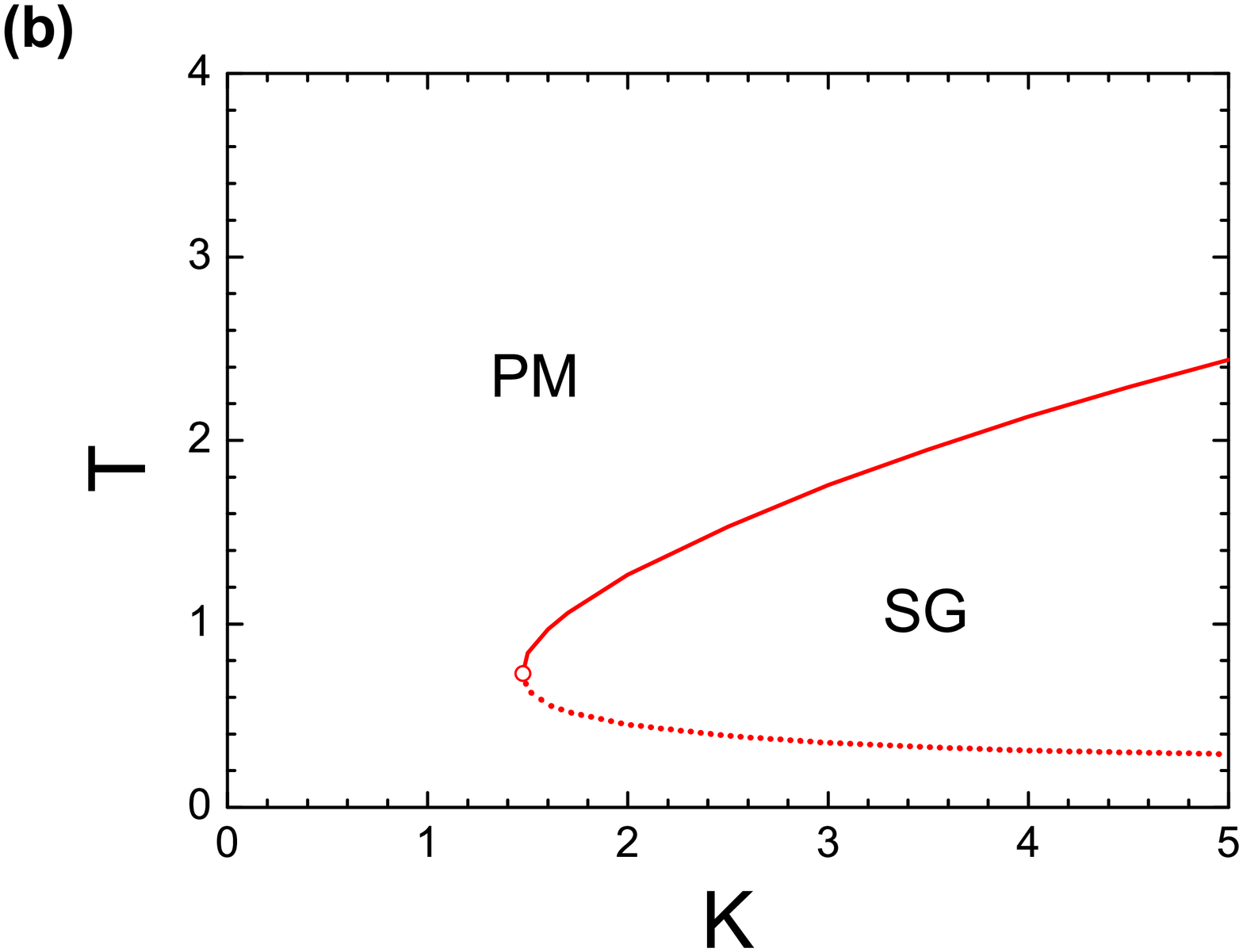}
\includegraphics[width=0.43\textwidth]{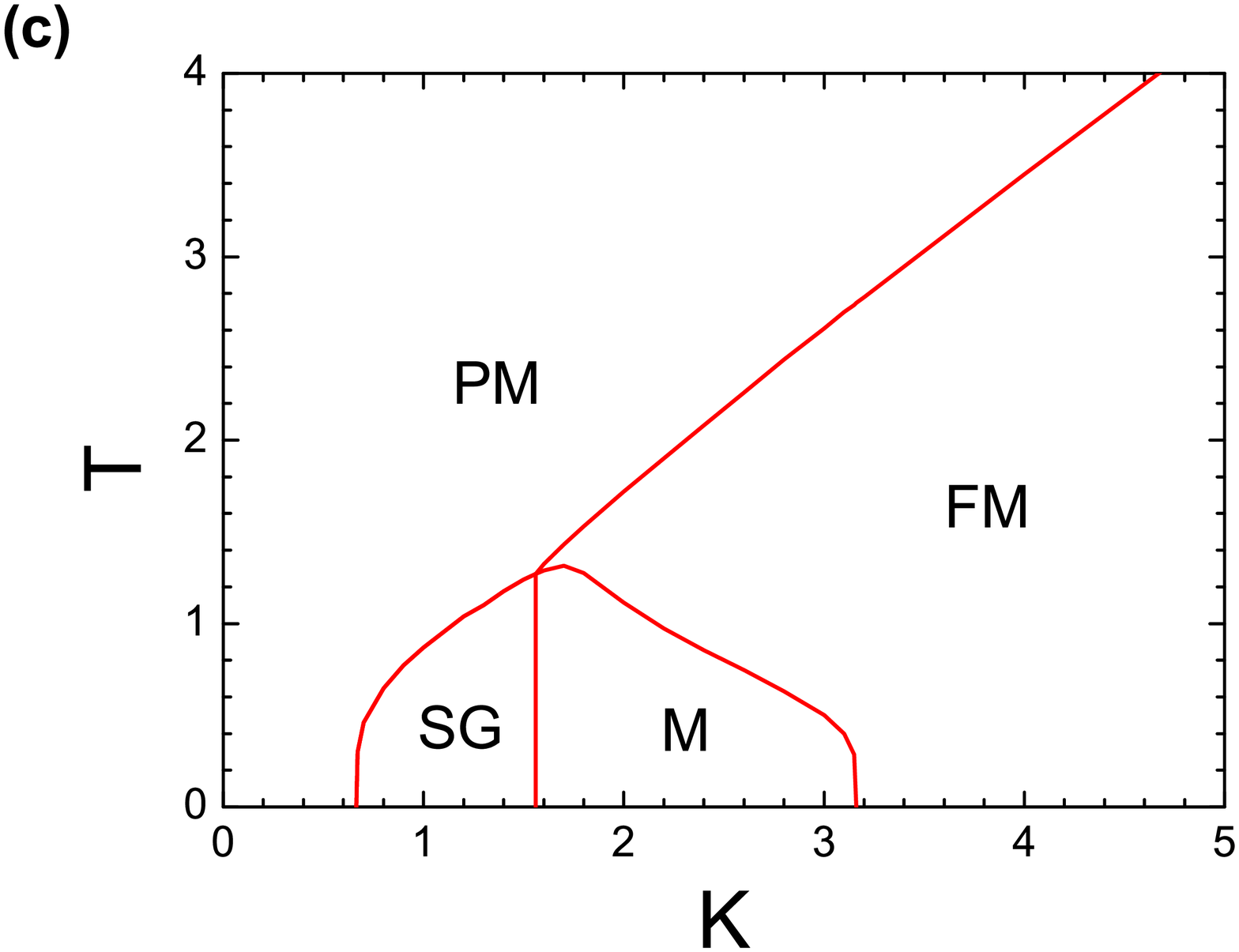}
\includegraphics[width=0.43\textwidth]{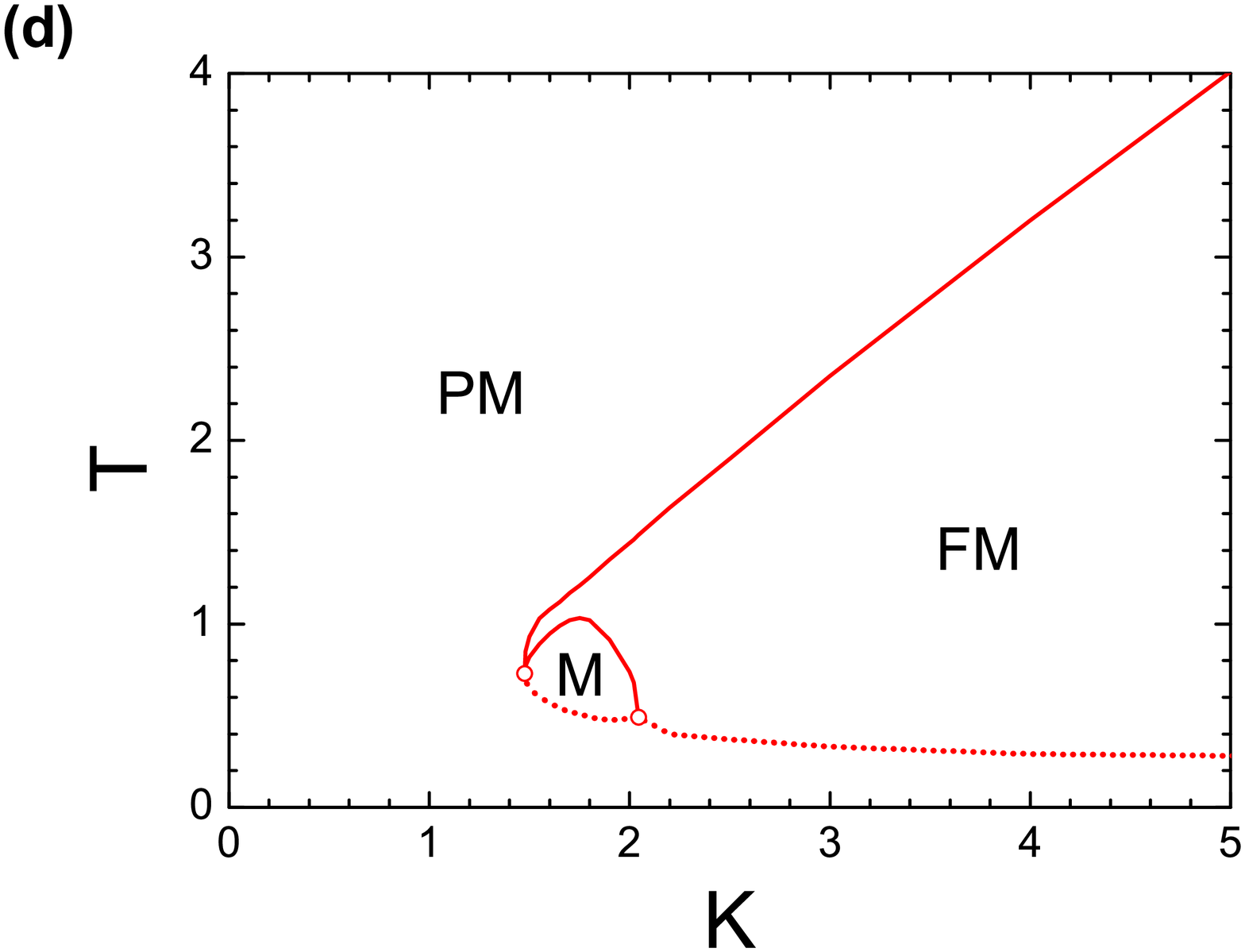}
\caption{(Color online) The $T-K$ phase diagrams for $\rho = 0.5$ and $D=0.0$ (a),  $\rho = 0.5$ and $D=1.0$ (b), $\rho = 0.75$ and $D=0.0$ (c), and $\rho = 0.75$ and $D=1.0$ (d).  Here, $N = 100$ and $\lambda = 2.5$. }\label{4}
\end{figure}

The graphs in Fig. 4 show the $T-K$ phase diagrams obtained for $\lambda=2.5$.
As mean degree $K$ increases, degrees of many vertices become larger. As a result, the number of hubs also increases.  Therefore, $K$ plays a fundamental role in the appearance of ordered phases like SG, FM, and M at nonzero $T$.
As shown in Fig. 4(a), in the case of $D=0.0$, the PM-SG transition occurs for $K$ values greater than a specific threshold ($K_{t} = 0.665$), and there is no inverse transition. When $D$ becomes nonzero, as shown in Fig. 4(b), a TCP is located at a certain threshold value ($K_{t}=1.475$ for $D=1.0$). In addition, an inverse freezing, in the order of PM $\stackrel{\mathrm{2nd}}{\longrightarrow}$ SG $\stackrel{\mathrm{1st}}{\longrightarrow}$ PM, occurs for $K > K_{t}$  as the temperature is lowered. As $K$ gradually increases, $T_{g}$ also increases, however, $T_{p}$ decreases.  When only $\rho$ is increased under the same condition of Fig. 4(a), as shown in Fig. 4(c), FM and M phases as well as PM and SG phases occur, but no inverse transition exists.  In Fig. 4(c), the location of the multicritical point is (1.560, 1.270).

When $D$ is increased under the condition of Fig. 4(c), a very peculiar inverse transition occurs, as shown in Fig. 4(d). In Fig. 4(d), no SG phase exists, but there is a complex inverse transition, in the order of PM $\stackrel{\mathrm{2nd}}{\longrightarrow}$ FM $\stackrel{\mathrm{2nd}}{\longrightarrow}$ M $\stackrel{\mathrm{1st}}{\longrightarrow}$ PM,  which occurs in the range of $1.475 \leq K \leq 2.045$, as the temperature is reduced.
 Furthermore, there are two TCPs at $K = 1.475$ and $K=2.045$. It is uncertain whether this inverse transition can be called an inverse freezing or an inverse melting. In addition, an inverse transition, in the order of PM $\stackrel{\mathrm{2nd}}{\longrightarrow}$ FM $\stackrel{\mathrm{1st}}{\longrightarrow}$ PM, occurs for $K > 2.045$,  as the temperature is reduced. This inverse transition corresponds to an inverse melting, according to the definition of Schupper and Shnerb \cite{Schupper04}. Therefore, we checked that two alternative types of inverse transition, in addition to inverse freezing, can be discovered by controlling $K$ using proper values of $\rho$ and $D$.

\begin{figure}
\includegraphics[width=0.43\textwidth]{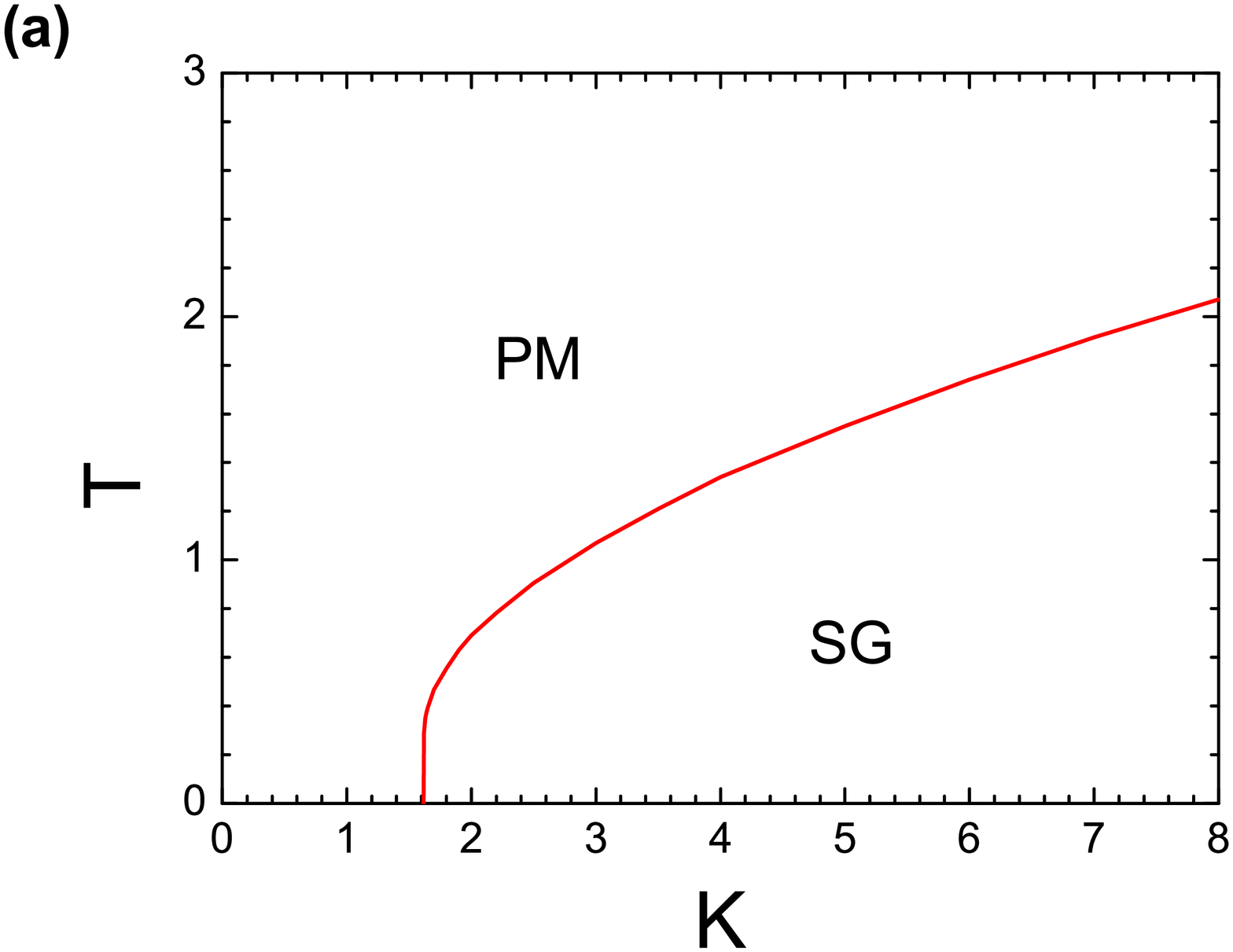}
\includegraphics[width=0.43\textwidth]{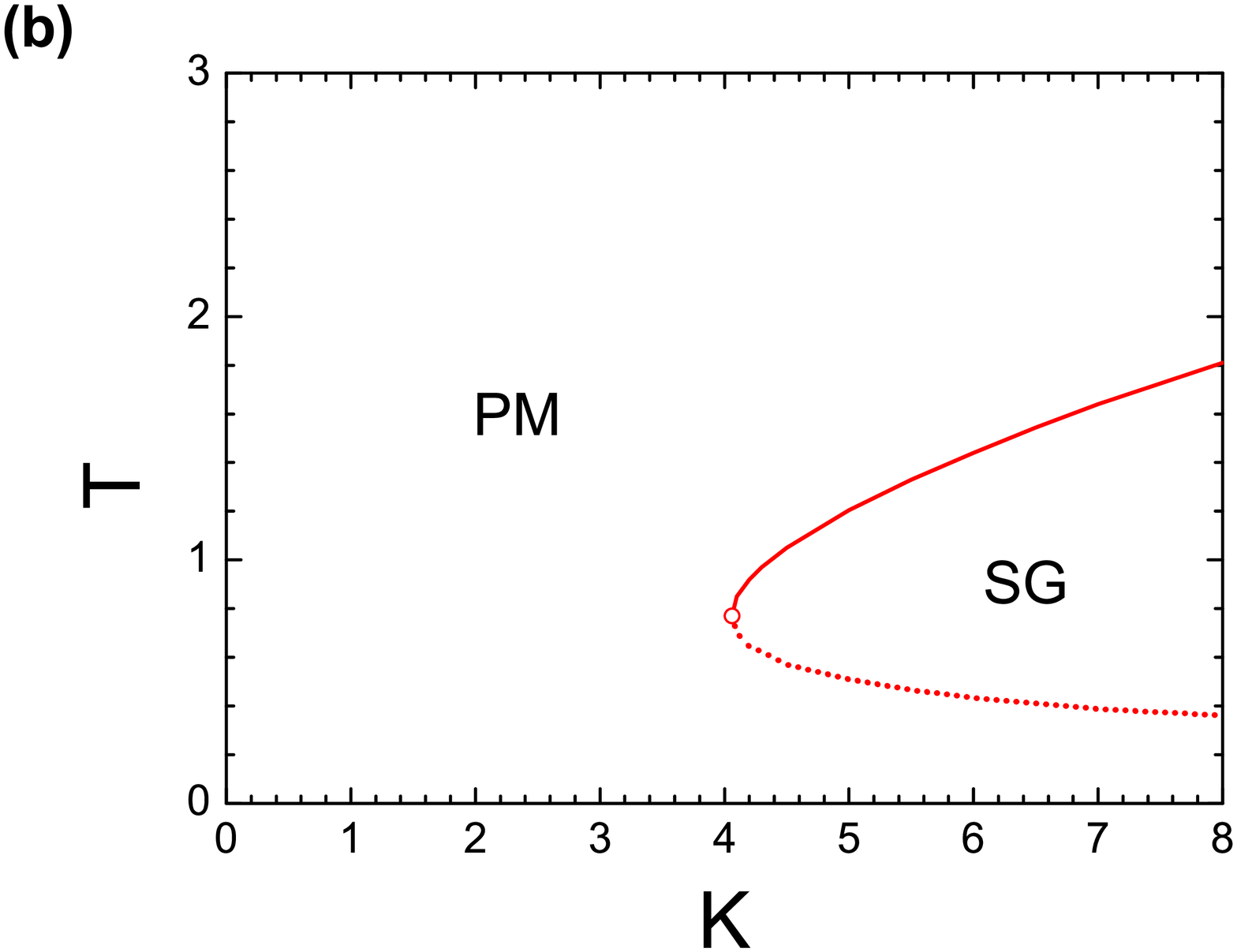}
\includegraphics[width=0.43\textwidth]{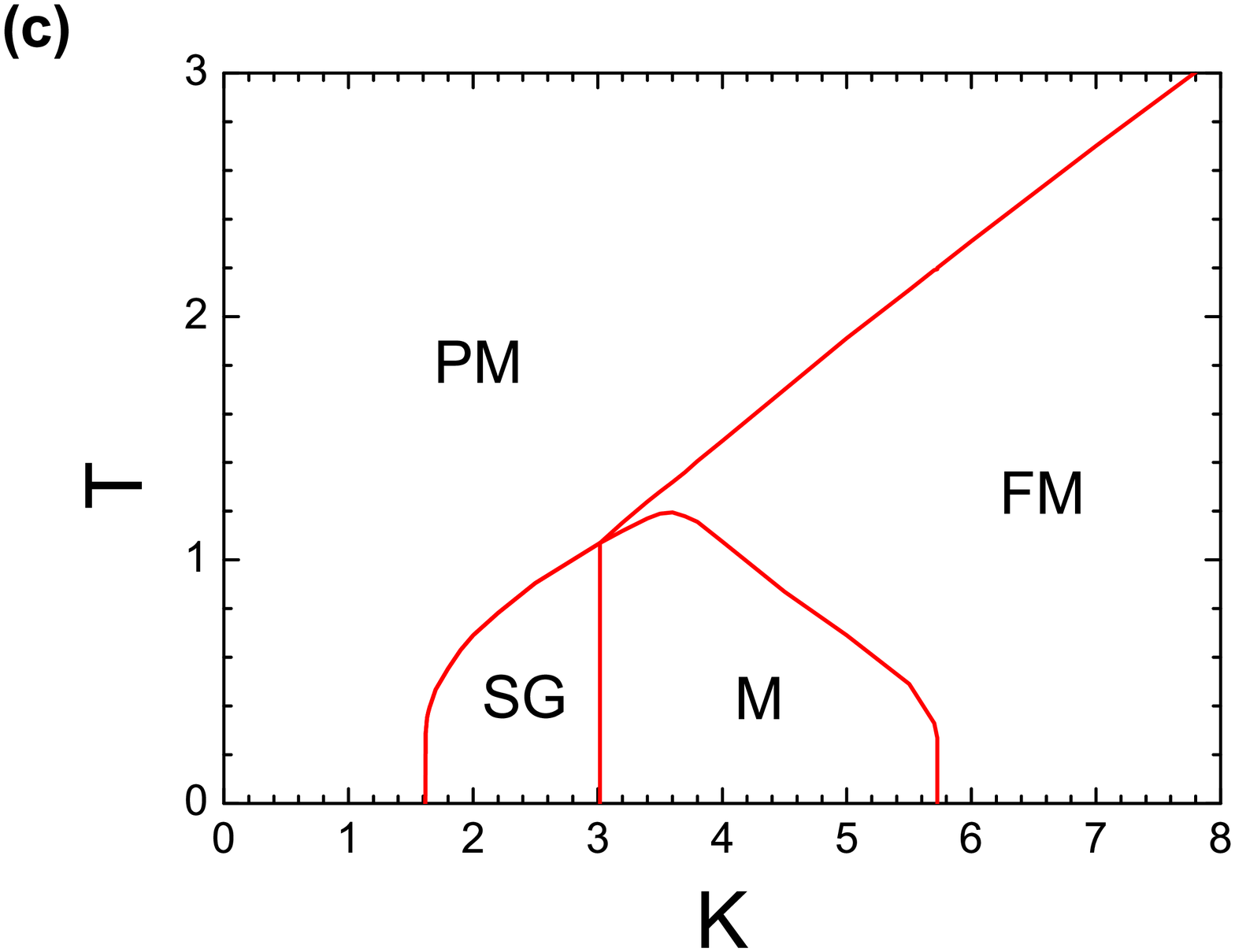}
\includegraphics[width=0.43\textwidth]{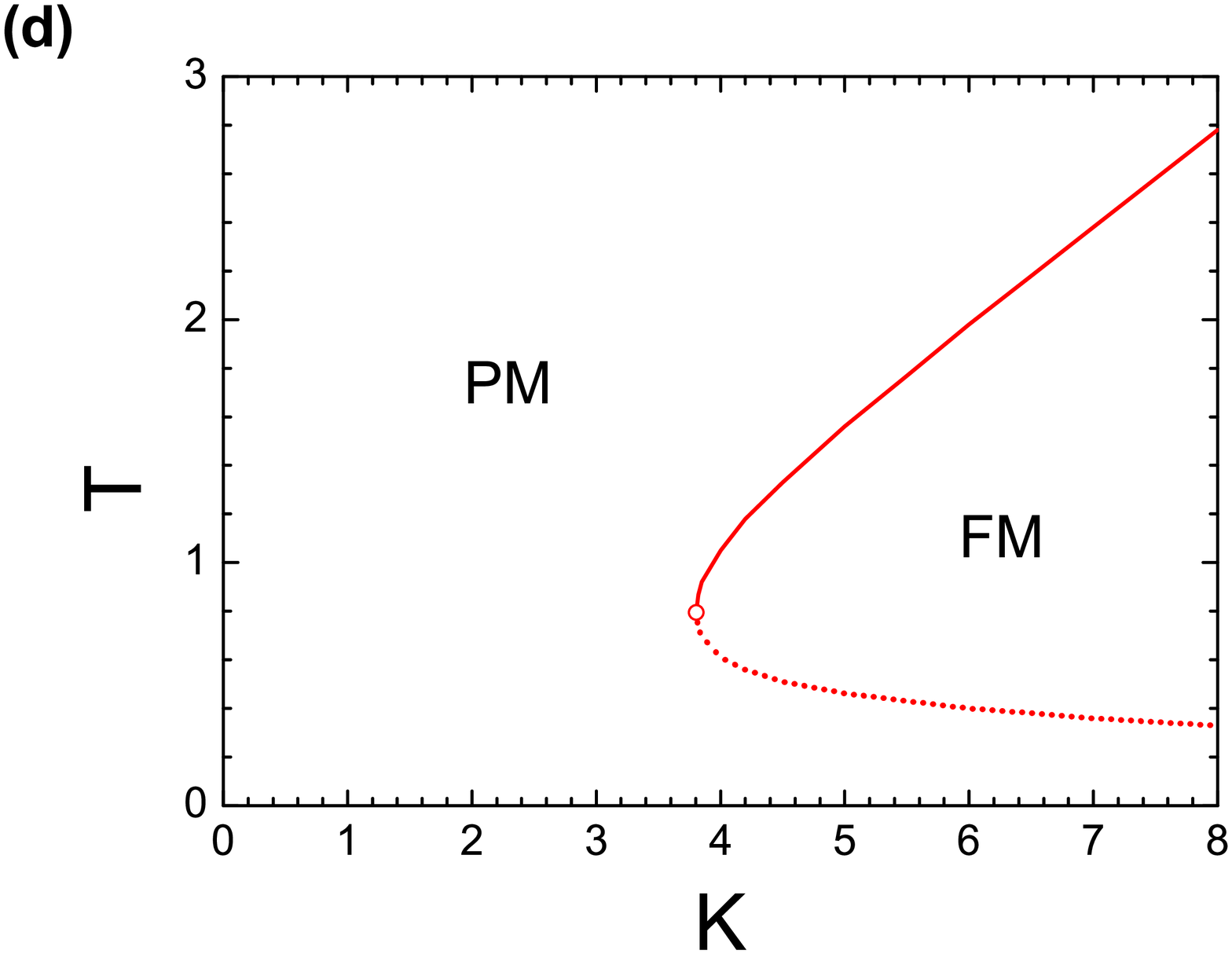}
\caption{(Color online) The $T-K$ phase diagrams for $\rho = 0.5$ and $D=0.0$ (a),  $\rho = 0.5$ and $D=1.0$ (b), $\rho = 0.75$ and $D=0.0$ (c), and $\rho = 0.75$ and $D=1.0$ (d).  Here, $N = 100$ and $\lambda = 6.0$. }\label{5}
\end{figure}

The graphs in Fig. 5 show the $T-K$ phase diagrams obtained for $\lambda=6.0$, which are given for comparison with Fig. 4. Figure 5(a) shows a similar result to Fig. 4(a), except for the threshold value ($K_{t} = 1.615$). Figure 5(b) is also similar to Fig. 4(b), but the threshold value, at which a TCP is located, is increased ($K_{t} = 4.060$). While Fig. 5(c) is also similar to Fig. 4(c), phase boundaries are shifted to a larger value of $K$. The location of the multicritical point is also shifted to (3.020, 1.070). In Fig. 5(d), the M phase disappears for $\lambda=6.0$ and an inverse melting, in the order of PM $\stackrel{\mathrm{2nd}}{\longrightarrow}$ FM $\stackrel{\mathrm{1st}}{\longrightarrow}$ PM, occurs for $K > 3.805$,  as the temperature is reduced.

\begin{figure}
\includegraphics[width=0.43\textwidth]{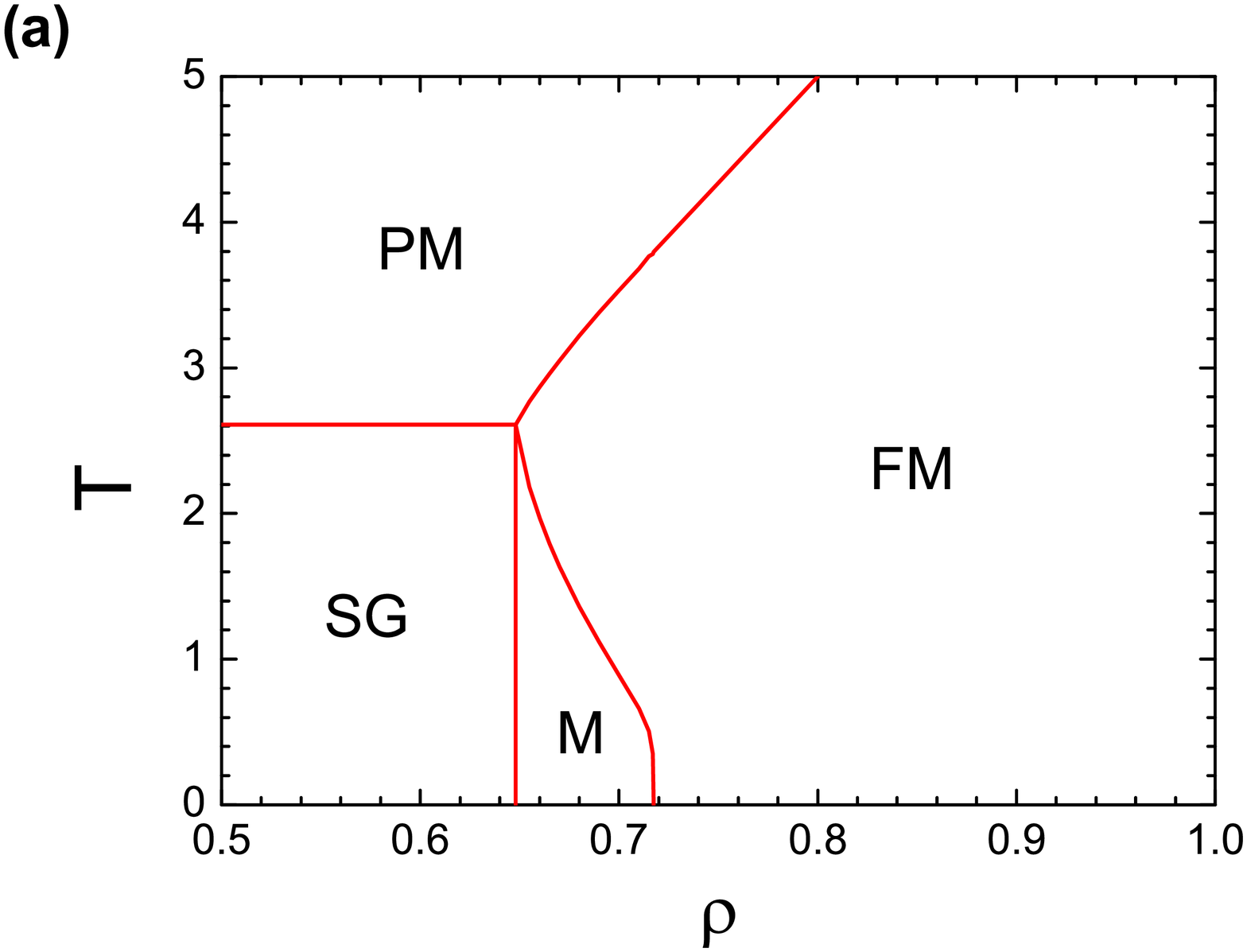}
\includegraphics[width=0.43\textwidth]{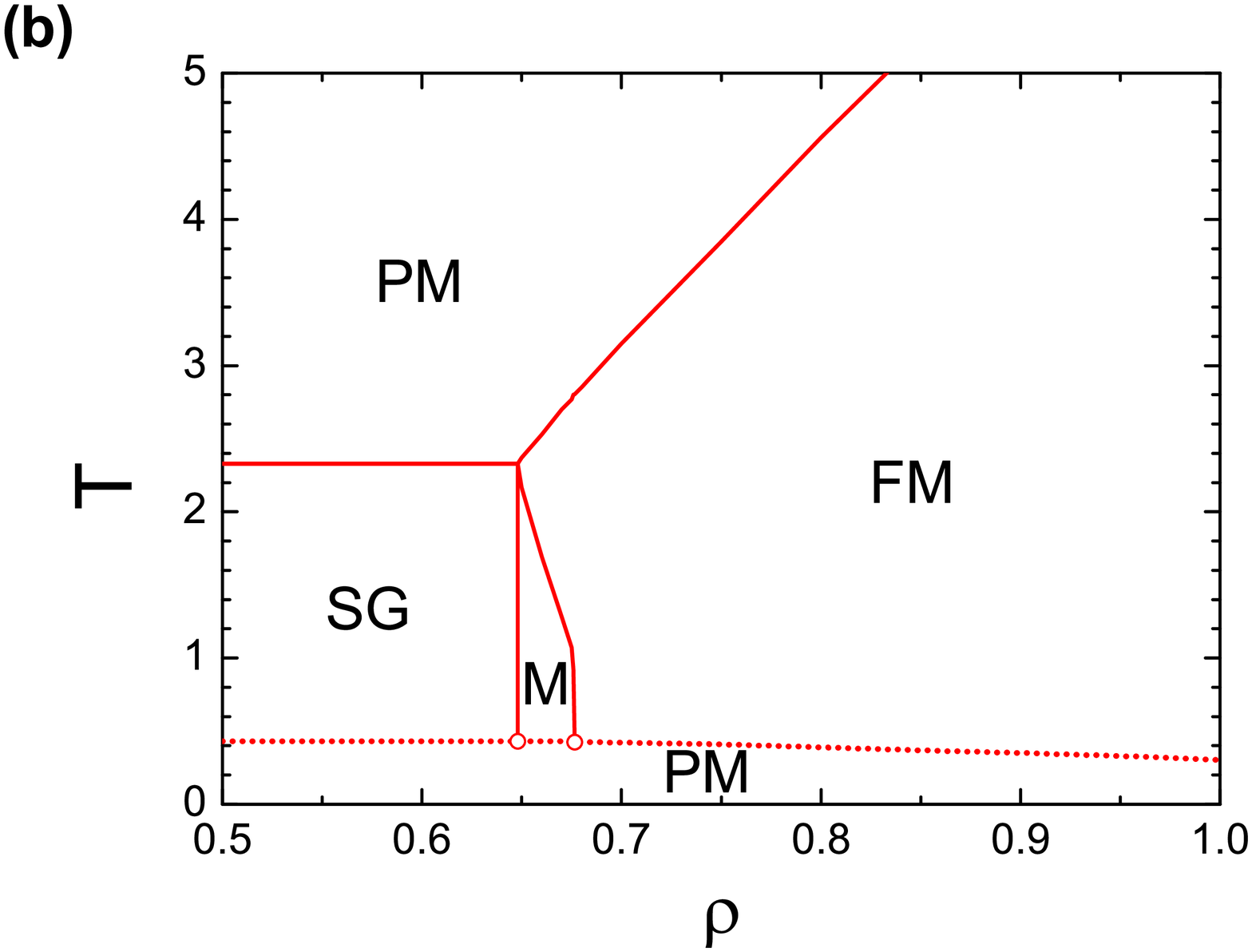}
\includegraphics[width=0.43\textwidth]{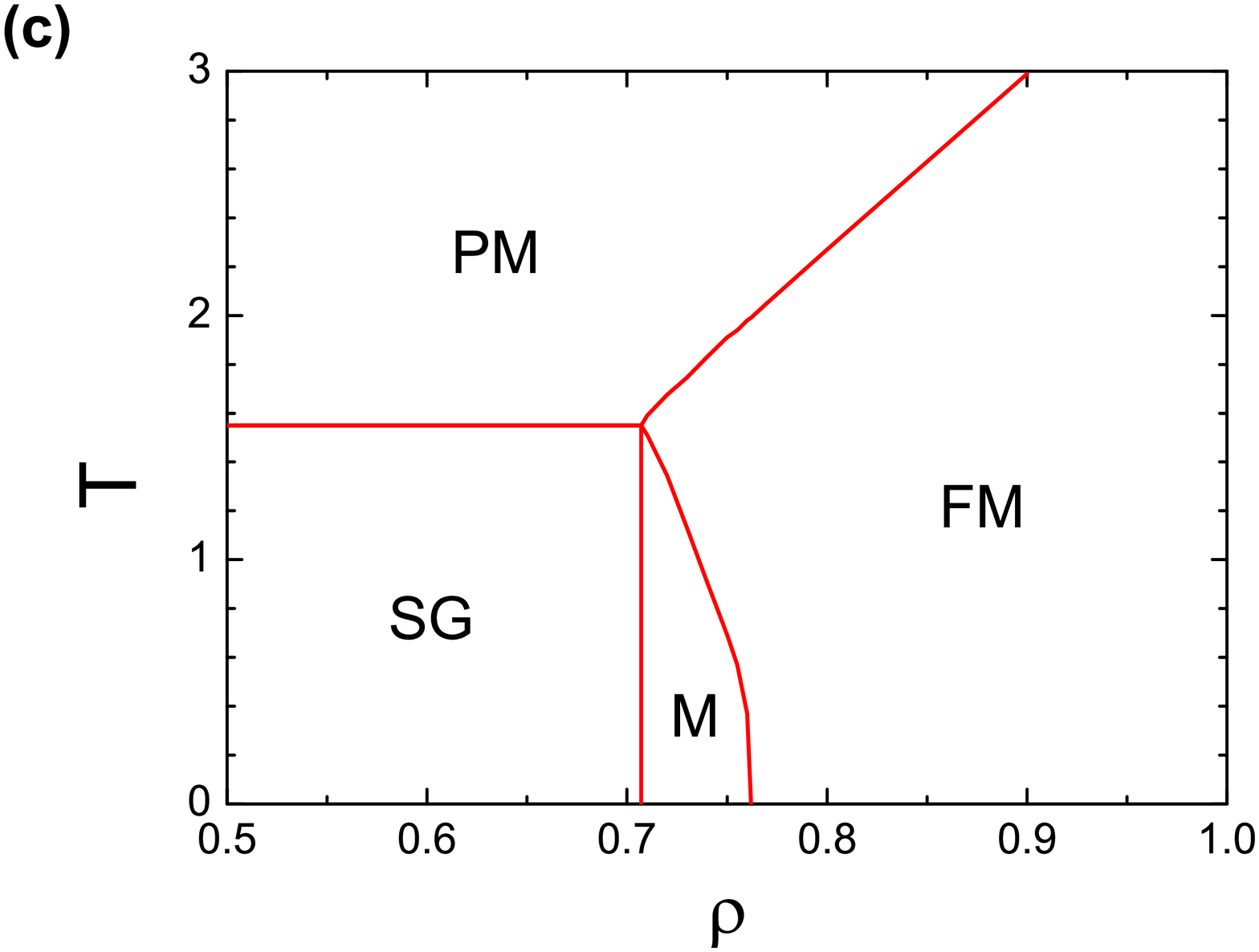}
\includegraphics[width=0.43\textwidth]{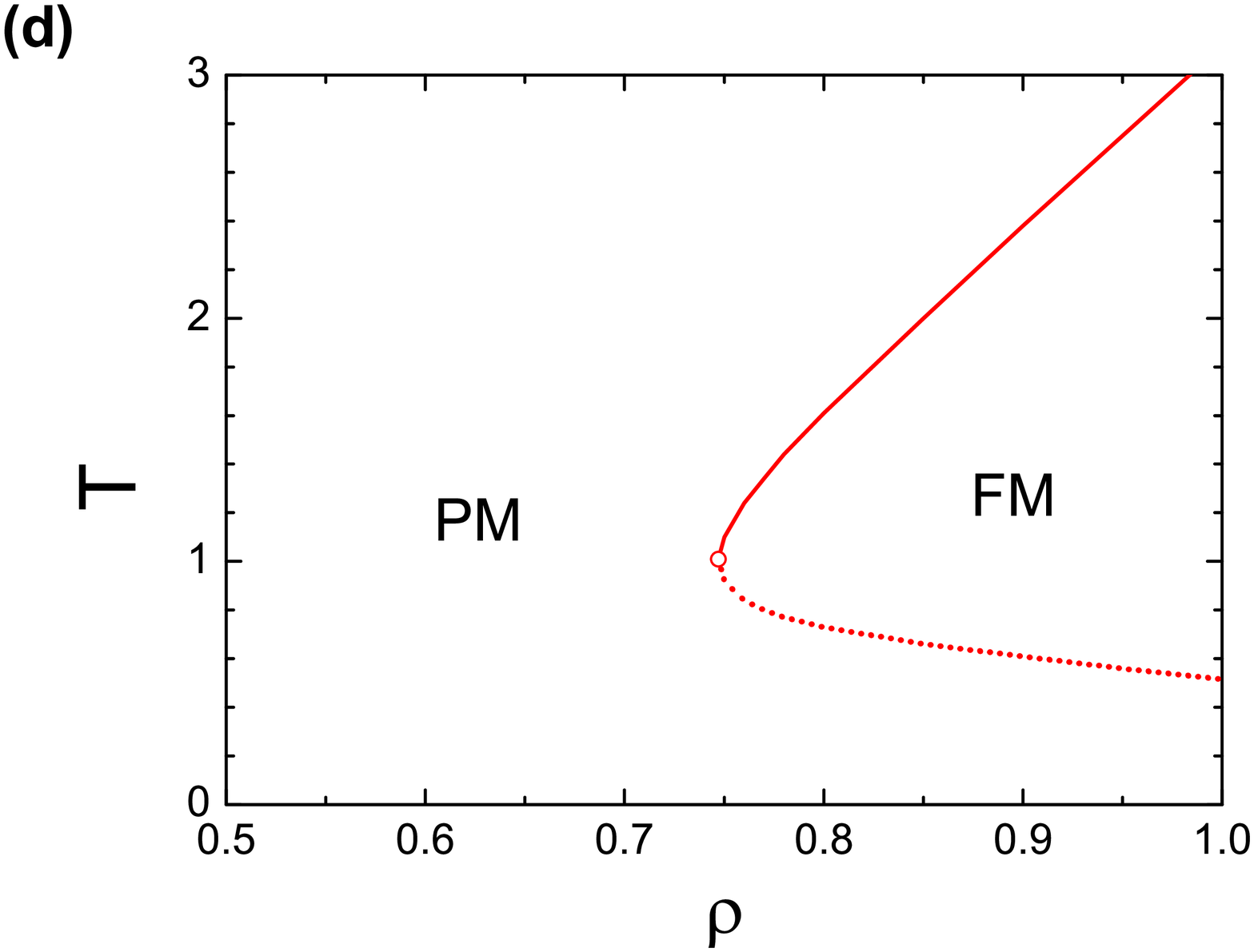}
\caption{(Color online) The $T-\rho$ phase diagrams for $\lambda = 2.5$ and $D=0.0$ (a),  $\lambda = 2.5$ and $D=1.5$ (b), $\lambda = 6.0$ and $D=0.0$ (c), and $\lambda = 6.0$ and $D=1.5$ (d).  Here, $N = 100$ and $K = 5.0$. }\label{6}
\end{figure}

 Figure 6 shows the $T-\rho$ phase diagrams obtained for several conditions. Since $\rho$ has a value between fully frustrated case ($\rho = 1/2$) and purely ferromagnetic one ($\rho = 1$), the increase of $\rho$ plays a role in replacing SG phase by FM or M phase.
 In Fig. 6(a) with $\lambda = 2.5$ and $D=0.0$, the PM-SG transition occurs when $\rho$ is below 0.648. In the region of $0.648 < \rho < 0.718$, successive transitions occur in the order of PM $\stackrel{\mathrm{2nd}}{\longrightarrow}$ FM $\stackrel{\mathrm{2nd}}{\longrightarrow}$ M, as the temperature is lowered. For $\rho \geq 0.718$, there is only the PM-FM transition.  In Fig. 6(a), the location of the multicritical point is (0.648, 2.610). When $D$ is increased to 1.5, as shown in Fig. 6(b), the inverse transitions occur throughout the range of $\rho$. When $\rho$ is smaller than 0.648, inverse freezing occurs in the order of PM $\stackrel{\mathrm{2nd}}{\longrightarrow}$ SG $\stackrel{\mathrm{1st}}{\longrightarrow}$ PM,  as the temperature is reduced. As can be checked through the comparison between Figs. 6(a) and 6(b), the value $\rho=0.648$, the location of the Toulouse vertical line determining the SG-M transition does not depend on the $D$ value. Instead, the location of the multicritical point is shifted to (0.648, 2.330) with $D=1.5$. However, for $0.648 < \rho < 0.677$, a complex inverse transition occurs  in the order of PM $\stackrel{\mathrm{2nd}}{\longrightarrow}$ FM $\stackrel{\mathrm{2nd}}{\longrightarrow}$ M $\stackrel{\mathrm{1st}}{\longrightarrow}$ PM. Moreover, there are two TCPs at $\rho = 0.648$ and $0.677$. For $\rho \geq 0.677$, an inverse melting occurs in the order of PM $\stackrel{\mathrm{2nd}}{\longrightarrow}$ FM $\stackrel{\mathrm{1st}}{\longrightarrow}$ PM,  as the temperature is lowered. Therefore, for sufficiently large $D$, we can find not only inverse freezing but also other inverse transitions by properly controlling the value of $\rho$.

 When only $\lambda$ is increased under the same condition of Fig. 6(a), Fig. 6(c) represents a similar result to Fig. 6(a), but $T_{g}$ is lowered and the location of the Toulouse line is shifted in the direction of large $\rho$. The location of the multicritical point is thus shifted to (0.707, 1.550). When $D$ is increased under the condition of Fig. 6(c), only PM and FM phases remain. The TCP is located at $\rho = 0.75$, and for larger $\rho$ values, inverse melting occurs in the order of PM $\stackrel{\mathrm{2nd}}{\longrightarrow}$ FM $\stackrel{\mathrm{1st}}{\longrightarrow}$ PM,  as the temperature is reduced.

 \begin{figure}
\includegraphics[width=0.43\textwidth]{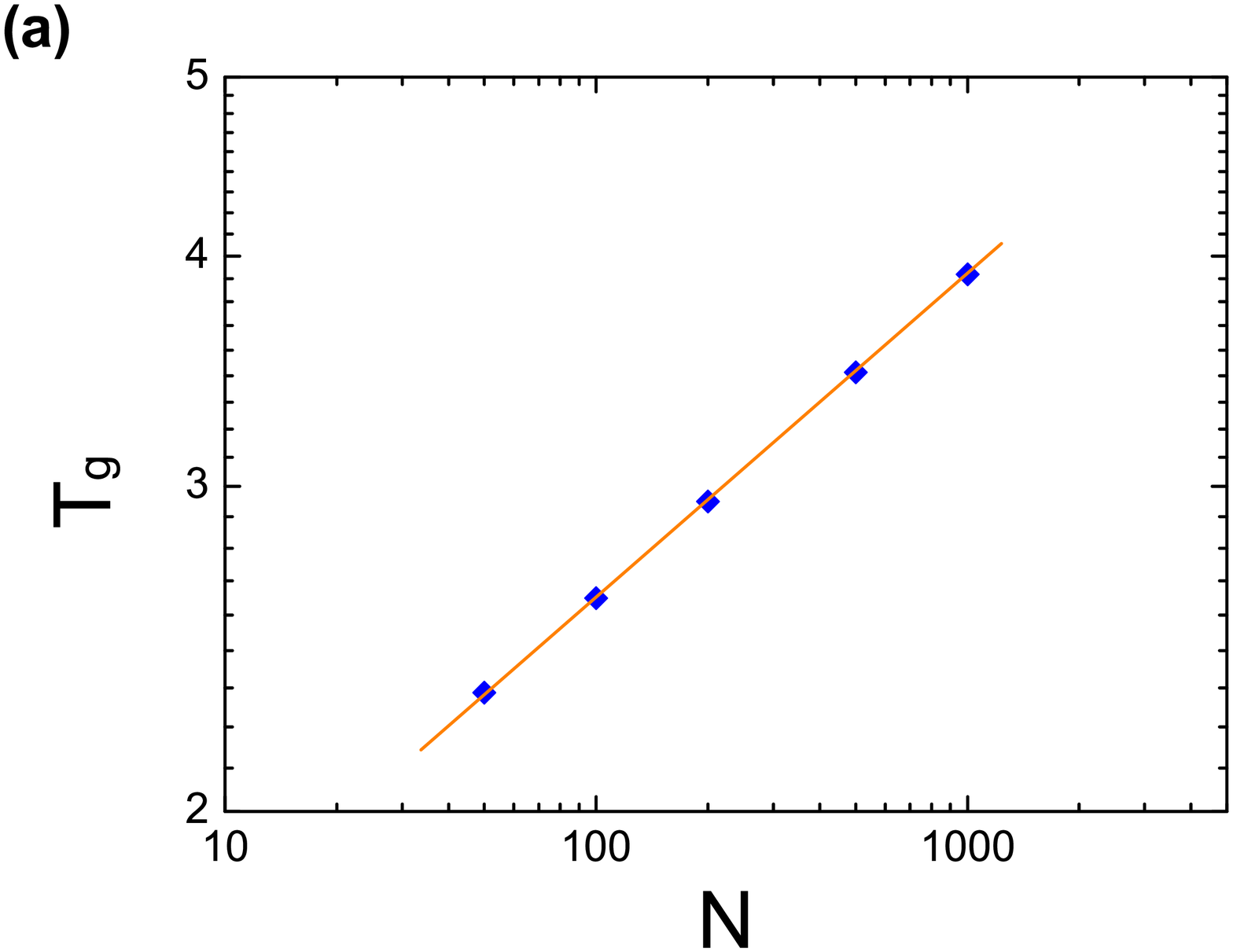}
\includegraphics[width=0.43\textwidth]{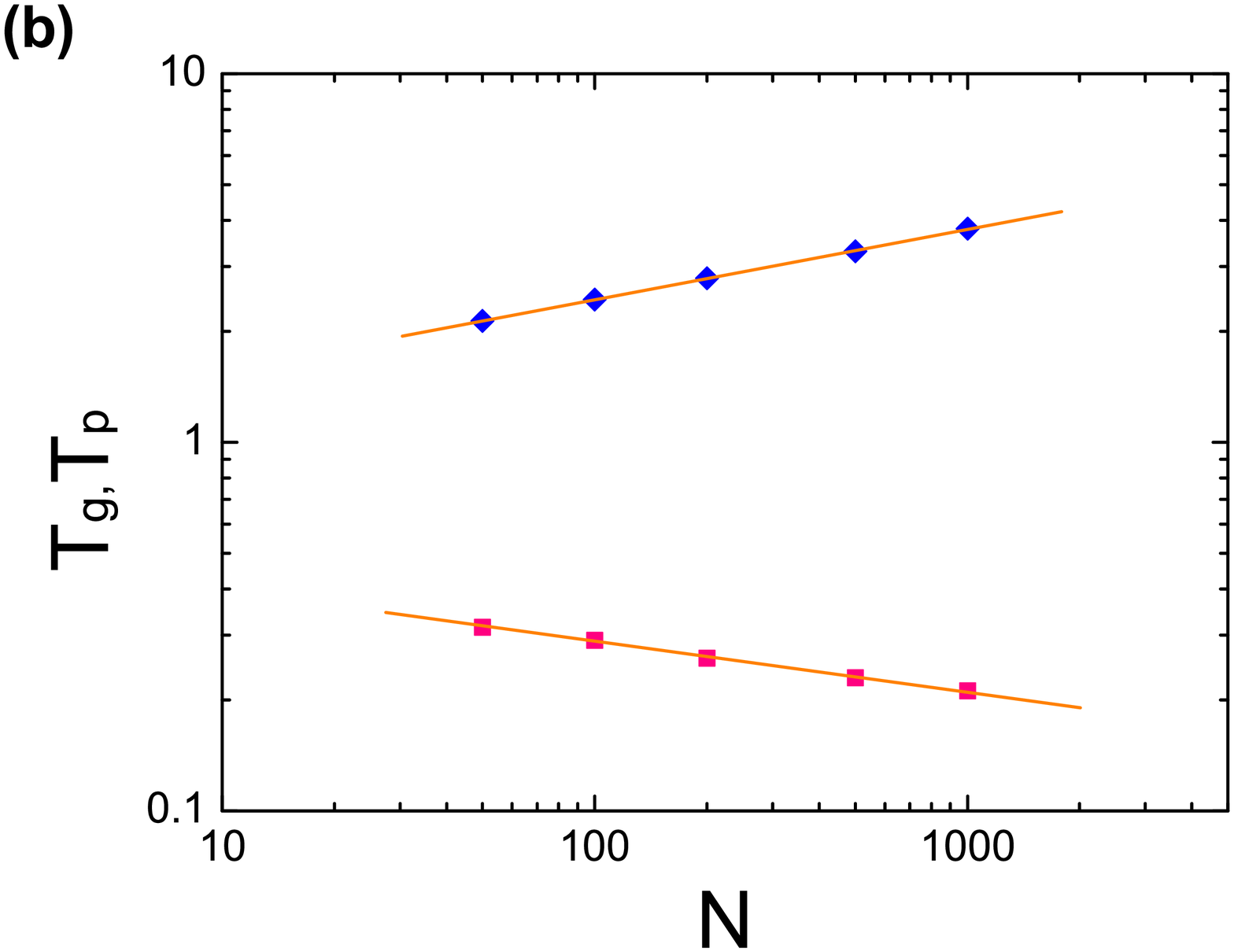}
\includegraphics[width=0.43\textwidth]{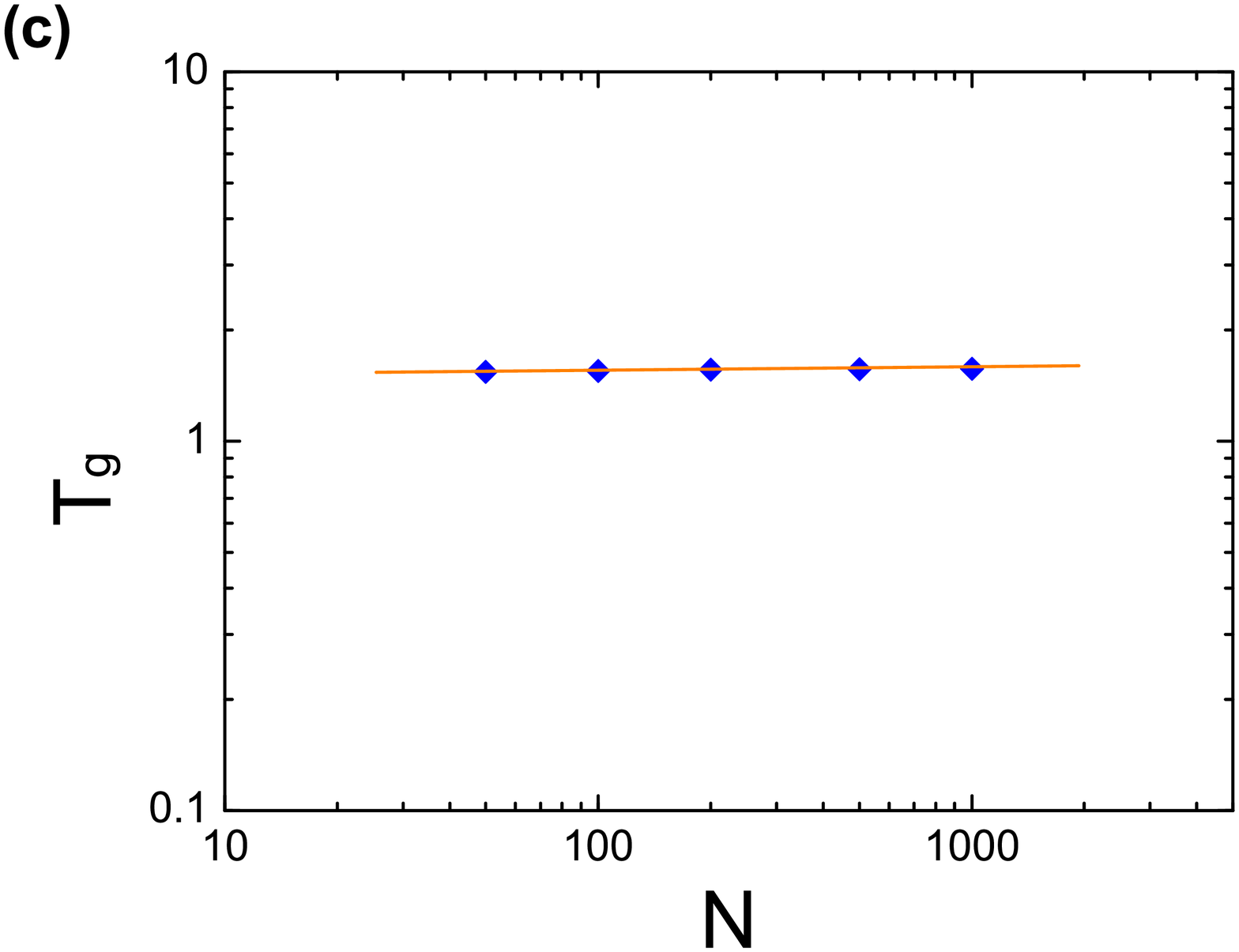}
\includegraphics[width=0.43\textwidth]{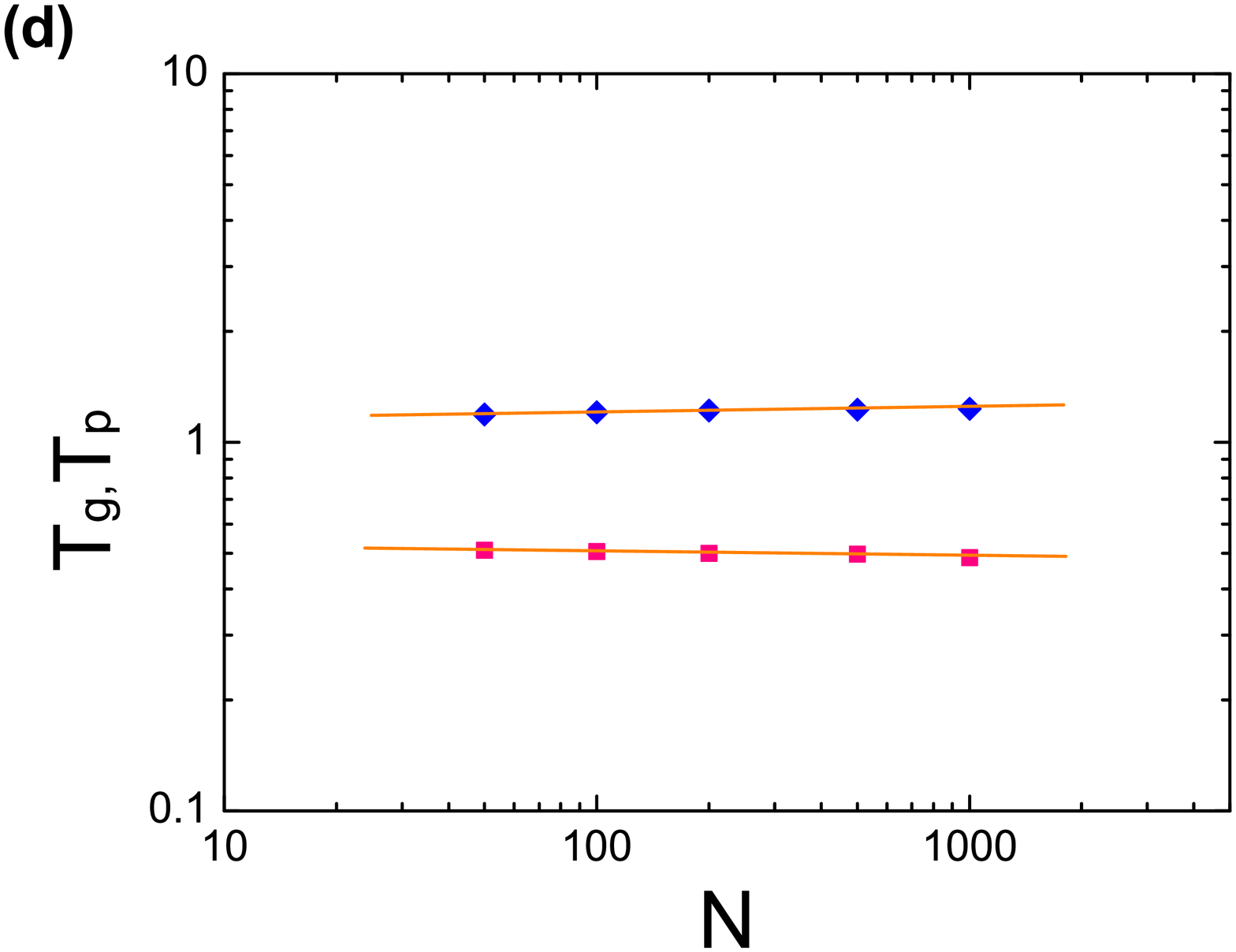}
\caption{(Color online) $T_{g}$($T_{p}$) vs. $N$ at $\lambda = 2.5$ and $D=0.0$ (a),  $\lambda = 2.5$ and $D=1.0$ (b), $\lambda = 6.0$ and $D=0.0$ (c), and $\lambda = 6.0$ and $D=1.0$ (d).  Here, $\rho = 0.5$ and $K = 5.0$. In the figures, the blue diamonds and red rectangles indicate $T_{g}$ and $T_{p}$, respectively.}\label{7}
\end{figure}

All the figures we have checked were drawn for $N=100$. However, our previous theoretical results were obtained under the assumption of $N \to \infty$. Therefore, we should agree that the value of $N=100$ is small when compared with the assumption. Then, we must investigate whether such referred characteristics of each figure are still satisfied, even for larger $N$ values than 100. Figure 7 shows the dependence of $T_{g}$ and $T_{p}$ on $N$.

Figure 7(a) clearly shows that $T_{g}$ depends on $N$ for $\lambda = 2.5$. According to fitting, we check that $\log T_{g} \simeq 0.174 \log N + 0.068$. In Fig.7(b) with $D=1.0$, inverse freezing by $T_{g}$ and $T_{p}$ exists, satisfying the fitting of $\log T_{g} \simeq 0.191 \log N + 0.004$ and $\log T_{p} \simeq -0.135 \log N - 0.272$, respectively. However, for $\lambda = 6.0$, $T_{g}$ and $T_{p}$ show little dependence on $N$, irrespective of $D$ values [$D=0.0$ in Fig. 7(c) and $D=1.0$ in Fig. 7(d)]. Such little dependence seems to be natural, because $M$, $R$, and $Q$ are independent of $N$ in the ER case ($p_{i} = 1/N$). Therefore, Fig. 7 represents all the results of previous figures that are quantitatively changed by the increase of the $N$ value, i.e., $T_{g}$ increases but $T_{p}$ decreases by an increase of $N$, and the changes become more sensitive as  $\lambda$ approaches 2.0. Since we can obtain the quantitative relation between $T_{g}$ (or other transition temperatures) and $N$ through the fitting method used in Fig. 7, we can draw the phase diagrams for all possible $N$ networks. Therefore, we can say that our previous results obtained from $N=100$ are still practically useful, even for the analysis of networks with very large $N$.

\section{Conclusions}

We have studied the inverse transitions on SF networks through the static model. As already proven in Ref. \cite{Kim05}, the
static model enables one to study the SG problem by generalizing the dilute Ising spin-glass model with
infinite-range interactions to a model with inhomogeneous vertex weights. We could also obtain the phase diagrams consisting of PM,
FM, SG, and M phases, as functions of temperature $T$, the degree exponent
$\lambda$, the mean degree $K$, and the fraction of the ferromagnetic interactions $\rho$.
However, the present model is based on the GS model, which considers the three states, including the $S=0$ state, contrary to the Ising-based model, which considers only $S = \pm 1$. The $S=0$ state makes an essential feature when compared with the previous model \cite{Kim05}: $T_{g}$ has a finite value even when $2 < \lambda < 3$. This feature is opposite to the previous model, in which $T_{g}$ is infinite for $2 < \lambda < 3$.
Furthermore, when crystal field $D$ has nonzero values, the present model shows three types of inverse transitions under specific conditions of parameters $\lambda$, $\rho$, and $K$. The inverse freezing occurs in the order of PM $\stackrel{\mathrm{2nd}}{\longrightarrow}$ SG $\stackrel{\mathrm{1st}}{\longrightarrow}$ PM, as the temperature is reduced. The inverse melting occurs in the order of PM $\stackrel{\mathrm{2nd}}{\longrightarrow}$ FM $\stackrel{\mathrm{1st}}{\longrightarrow}$ PM. The third case can also occur in the order of PM $\stackrel{\mathrm{2nd}}{\longrightarrow}$ FM $\stackrel{\mathrm{2nd}}{\longrightarrow}$ M $\stackrel{\mathrm{1st}}{\longrightarrow}$ PM.
Therefore, these two main results of the present model have special features as follows: until now, there have been few network models which do not contain the divergence of  transitions  for $2 < \lambda < 3$. The present model may thus be considered as an original one to reveal critical behavior by including the third spin state ($S=0$). In addition, the present model has a merit in that three types of inverse transitions can be simultaneously investigated by one type of model.

Our investigations could be helpful in the understanding of various patterns in real systems with competing interactions, such as social networks.
Usually, Ising-like two-spin states may be considered to represent two different opinions in a
society. In a similar way, we can apply the present model with three-spin states to a real-world three-state systems. One good example is the presidential elections carried out in many countries. When there are two candidates running for election, one segment of the people vote for a candidate in the ruling party, while others vote for a candidate in the opposition party, and the remaining portion abstains from voting in the election. The existence of such abstentions is given by floating voters, who show frustration in deciding between the two candidates. Since SG is characterized by frustration, the SG phase can be a good analogy for a frustrated state in which the majority of people are floating voters, in spite of their firm will to vote.  Similarly, the PM phase can express a state with nearly the same number of supporters for the two candidates, the FM phase is a state with a superiority of one party, and the M phase is a complex state, including both a  little superiority of one party and floating voters. The result of the election, i.e., selection of the phase, can be determined by several external conditions, such as a degree of election fever ($T$), a degree of social mood toward indifference to the election ($D$), a relative approval rating by a public opinion poll ($\rho$), a degree exponent reflecting the network topology of the country holding the election ($\lambda$), and a mean degree of people in the country ($K$). When $D$ is zero, the SG phase by floating voters may be found even for the case of a very low election fever. However, when $D$ is nonzero, the PM phase made by people who are indifferent to the election can be found for low election fever. Further research based on the data of real votes may prove the usefulness of the present model.

We believe that the methods we used here can be applied to other spin systems placed on SF networks. Moreover, we expect that the present model will help us extend our perspective of SG systems and of inverse transitions not only in theoretical models but also in real-world systems.

\begin{acknowledgments}
The author thanks Jesuit Community colleagues at Sogang University for helpful comments. This work was supported by the Formation Fund for Korean Jesuit Scholastics.
\end{acknowledgments}

\end{document}